\documentclass[11pt]{article}
\usepackage{fullpage}
\usepackage{times}
\usepackage{graphicx}
\usepackage{amsmath,amssymb,amstext}

\raggedright

\begin{document}
  \title{Simulation of thin hyperelastic shells with the Material Point Method}
  \author{Biswajit Banerjee
          \footnote{E-mail: banerjee@eng.utah.edu.
          Phone: (801) 585-5239 Fax: (801) 585-0039}
          \\
          Department of Mechanical Engineering, University of Utah, 
          Salt Lake City, UT 84112, USA}
  \maketitle
  \begin{abstract}
  A non-linear shell theory that includes transverse shear strains 
  and its implementation in the material point method framework are
  discussed.  The applicability of the shell implementation to model
  large deformations of thin shells is explored.  Results suggest that
  an implicit time stepping scheme may be required for improved 
  modeling of thin shells by the material point method.
  \end{abstract}

  \section{Shell Formulation}
  The continuum-based approach to shell theory has been chosen because
  of the relative ease of implementation of constitutive models in this
  approach compared to exact geometrical descriptions of the shell.  In 
  order to include transverse shear strains in the shell, a modified
  Reissner-Mindlin assumption is used.  The major assumptions of the
  shell formulation are~\cite{Belyt00,Lewis98}
  \begin{enumerate}
    \item The normal to the mid-surface of the shell remains straight
          but not necessarily normal.  The direction of the initial
          normal is called the ``fiber'' direction and it is the evolution
          of the fiber that is tracked.
    \item The stress normal to the mid-surface vanishes (plane stress)
    \item The momentum due to the extension of the fiber and the momentum
          balance in the direction of the fiber are neglected.
    \item The curvature of the shell at a material point is neglected.
  \end{enumerate}
  The shell formulation is based on a plate formulation by Lewis et al.
  ~\cite{Lewis98}.  A discussion of the formulation follows.  

  The velocity field in the shell is given by
  \begin{equation}
    \mathbf{w}(\alpha,\beta) = \mathbf{u}(\alpha,\beta) 
        +z~\boldsymbol{\omega}(\alpha,\beta)\times \mathbf{n}(\alpha,\beta)
        +\dot{z}~\mathbf{n}(\alpha,\beta) \label{eq:velField}
  \end{equation}
  where $\mathbf{w}$ is the velocity of a point in the shell,
        $\mathbf{u}$ is the velocity of the center of mass of the shell,
        $\mathbf{n}$ is the normal or director vector,
        $\boldsymbol{\omega}$ is the angular velocity of the director,
        $(\alpha,\beta)$ are orthogonal co-ordinates on the mid-surface of
        the shell, $z$ is the perpendicular distance from the mid-surface
        of the shell, and $\dot{z}$ is the rate of change of the length 
        of the shell director.

  Since momentum balance is not enforced for the motion in the direction
  of the director $\mathbf{n}$, the terms involving $\dot{z}$ are dropped
  in constructing the equations of motion.  These terms are also omitted
  in the deformation gradient calculation.  However, the thickness change
  in the shell is not neglected in the computation of internal forces and
  moments.  Equation~(\ref{eq:velField}) can therefore be written as
  \begin{equation}
    \mathbf{w}(\alpha,\beta) = \mathbf{u}(\alpha,\beta) 
        +z~\mathbf{r}(\alpha,\beta)
  \end{equation}
  where $\mathbf{r}$, the rotation rate of $\mathbf{n}$, is a vector
  that is perpendicular to $\mathbf{n}$.

  The velocity gradient tensor for $\mathbf{w}$ is used to compute the 
  stresses in the shell.
  If the curvature of the shell is neglected, i.e., the shell is piecewise
  plane, the velocity gradient tensor for $\mathbf{w}$ can be written
  as
  \begin{equation}
    \boldsymbol{\nabla}\mathbf{w} = \left[\boldsymbol{\nabla}^{(s)}\mathbf{u}+
        z~\boldsymbol{\nabla}^{(s)}\mathbf{r}\right] + 
        \mathbf{r}\otimes\mathbf{n}
     \label{eq:velGrad}
  \end{equation}
  where $\mathbf{r}\otimes\mathbf{n}$ represents the dyadic product, and 
  $\mathbf{\nabla}^{(s)}$ is the in-surface gradient operator, defined as,
  \begin{equation}
    \boldsymbol{\nabla}^{(s)} = \left[\boldsymbol{\nabla}(~~)\right]
                            \bullet\mathbf{I}^{(s)}~.
  \end{equation}
  The $\bullet$ represents a tensor inner product and $\mathbf{I}^{(s)}$
  is the in-surface identity tensor (or the projection operator), defined as,
  \begin{equation}
    \mathbf{I}^{(s)} = \mathbf{I} - \mathbf{n}\otimes\mathbf{n}.
  \end{equation}
  It should be noted that, for accuracy, the vector $\mathbf{n}$ should not
  deviate significantly from the actual normal to the surface (i.e., the
  transverse shear strains should be small).

  The determination of the shell velocity tensor 
  $\boldsymbol{\nabla}\mathbf{w}$ requires the determination of the 
  center of mass velocity $\mathbf{u}$ of the shell.  This quantity is
  determined using the balance of linear momentum in the shell.
  The local three-dimensional equation of motion for the shell is, in the
  absence of body forces,
  \begin{equation}
    \boldsymbol{\nabla}\bullet\boldsymbol{\sigma} = \rho~\mathbf{a}
    \label{eq:momBal3D}
  \end{equation}
  where $\mathbf{sigma}$ is the stress tensor, $\rho$ is the density of
  the shell material, and $\mathbf{a}$ is the acceleration of the shell.
  The two-dimensional form of the linear momentum balance equation
  ~(\ref{eq:momBal3D}) with respect to the surface of the shell is
  given by
  \begin{equation}
    \boldsymbol{\nabla}^{(s)}\bullet \left<\boldsymbol{\sigma}\right> 
     = \rho~\mathbf{a}~.
    \label{eq:momBal2D}
  \end{equation}
  The acceleration of the material points in the shell are now due to
  the in-surface divergence of the average stress 
  $\left<\boldsymbol{\sigma}\right>$ in the shell, given by
  \begin{equation}
    \left<\boldsymbol{\sigma}\right> \colon= 
      \frac{1}{h}\int^{h^+}_{-h^-} \boldsymbol{\sigma}(z)~dz
    \label{eq:avStress}
  \end{equation}
  where $h^+$ is the ``thickness'' of the shell (along the director) from
  the center of mass to the ``top'' of the shell, $h^-$ is the thickness
  from the center of mass to the ``bottom'' of the shell, and $h = h^+ + h^-$.  
  The point of departure from the formulation of Lewis et al.~\cite{Lewis98} 
  is that instead of separate linear momentum balance laws for shell and 
  non-shell materials, a single global momentum balance is used and the 
  ``plane stress'' condition $\mathbf{\sigma}_{zz} = 0$ is enforced in the 
  shell stress update, where the subscript $zz$ represents the direction of 
  the shell director.  
  
  The shell director $\mathbf{n}$ and its rotation rate $\mathbf{r}$ also need
  to be known before the shell velocity gradient tensor 
  $\boldsymbol{\nabla}\mathbf{w}$ can be determined.  These quantities
  are determined using an equation for the conservation of angular 
  momentum~\cite{Schreyer97}, given by
  \begin{equation}
    \boldsymbol{\nabla}^{(s)}\bullet\mathbf{M} - 
      \mathbf{n}\bullet\left<\boldsymbol{\sigma}\right>\bullet\mathbf{I}^{(s)}
    = \frac{1}{12}~\rho~h^2~\dot{\mathbf{r}}
    \label{eq:rotInertia}
  \end{equation}
  where $\dot{\mathbf{r}}$ is the rotational acceleration of $\mathbf{n}$,
  $\rho$ is the density of the shell material, and $\mathbf{M}$ is the 
  average moment, defined as
  \begin{equation}
    \mathbf{M} \colon= \mathbf{I}^{(s)}\bullet
                 \left[ \frac{1}{h}\int^{h^+}_{-h^-}\boldsymbol{\sigma}(z)~z~dz
                 \right]\bullet \mathbf{I}^{(s)}~.
    \label{eq:avMoment}
  \end{equation}

  The center-of-mass velocity $\mathbf{u}$, the director $\mathbf{n}$ and
  its rate of rotation $\mathbf{r}$ provide a means to obtain the velocity
  of material points on the shell.  The shell is divided into a number
  of layers with discrete values of $z$ and the layer-wise gradient of the 
  shell velocity is used to compute the stress and deformation in each layer 
  of the shell.

  \section{Shell Implementation for the Material Point Method}
  The shell description given in the previous section has been implemented
  such that the standard steps of the material point method~\cite{Sulsky94}
  remain the same for all materials.  Some additional steps are performed for 
  shell materials.  These steps are encapsulated within the shell
  constitutive model.  

  The steps involved for each time increment $\Delta t$ are discussed below.
  The superscript $n$ represents the value of the state variables at time
  $n~\Delta t$ while the superscript $n+1$ represents the value at time
  $(n+1)~\Delta t$.  Note that $\Delta t$ need not necessarily be constant.
  In the following, the subscript $p$ is used to index material point variables
  while the subscript $v$ is used to index grid vertex variables.  The notation
  $\sum_p$ denotes summation over material points and $\sum_v$ denotes
  summation over grid vertices.  Zeroth order interpolation functions 
  associated with each material point are denoted by $S^{(0)}_{p,v}$ while
  first order interpolation functions are denoted by $S^{(1)}_{p,v}$.
  \begin{enumerate}
    \item  {\bf Interpolate state data from material points to 
      the grid.}\\
      The state variables are interpolated from the material points to
      the grid vertices using the contiguous generalized interpolation
      material point (GIMP) method~\cite{Bard00a}.  In the GIMP method
      material points are defined by particle characteristic functions
      $\chi_p(\mathbf{x})$ which are required to be a partition of unity,
      \begin{equation}
        \sum_p \chi_p (\mathbf{x}) = 1~~\forall~~\mathbf{x} \in \Omega
      \end{equation}
      where $\mathbf{x}$ is the position of a point in the body $\Omega$. 
      A continuous representation of the property $f(\mathbf{x})$ 
      is given by
      \begin{equation}      
        f(\mathbf{x}) = \sum_p f_p~\chi_p(\mathbf{x})~
      \end{equation}      
      where $f_p$ is the value at a material point.
      Similarly, a continuous representation of the grid data is given by
      \begin{equation}      
        g(\mathbf{x}) = \sum_v g_v~S_v(\mathbf{x})~
      \end{equation}      
      where
      \begin{equation} 
        \sum_v S_v(\mathbf{x}) = 1~~\forall~~\mathbf{x} \in \Omega~.
      \end{equation} 
      To interpolate particle data to the grid, the interpolation (or 
      weighting functions) $S^{(1)}_{p,v}$ are used, which are defined as
      \begin{equation}
        S^{(1)}_{p,v} = \frac{1}{V_p}\int_{\Omega_p\cap\Omega}
                            \chi_p(\mathbf{x})~S_v(\mathbf{x})~d{\mathbf{x}}~
      \end{equation}
      where $V_p$ is the volume associated with a material point, $\Omega_p$
      is the region of non-zero support for the material point, and
      \begin{equation} 
        \sum_v S^{(1)}_{p,v} = 1~~\forall~~\mathbf{x_p} \in \Omega_p~.
      \end{equation} 

      The state variables that are interpolated to the grid in this step
      are the mass ($m$), momentum ($m\mathbf{u}$), volume ($V$),
      external forces ($\mathbf{f}^{\text{ext}}$), temperature ($T$),
      and specific volume ($v$) using relations of the form
      \begin{equation}
        m_v = \sum_p m_p~S^{(1)}_{p,v}~.
      \end{equation}

      In our computations, bilinear hat functions $S_v$ were used that 
      lead to interpolation functions $S^{(1)}_{p,v}$ with non-zero support
      in adjacent grid cells and in the next nearest neighbor grid cells.
      Details of these functions can be found in reference~\cite{Bard00a}.

      For shell materials, an additional step is required to inhabit the 
      grid vertices with the interpolated normal rotation rate from the 
      particles.   However,
      instead of interpolating the angular momentum, the quantity 
      $ \mathbf{p}_p = m_p\mathbf{r}_p$ is interpolated to the grid using the 
      relation
      \begin{equation}
        \mathbf{p}_v = \sum_p \mathbf{p}_p~S^{(1)}_{p,v}~.
      \end{equation}
      At the grid, the rotation rate is recovered using
      \begin{equation}
        \mathbf{r}_v = \mathbf{p}_v / m_v
      \end{equation}
      This approximation is required because the moment of inertia contains
      $h^2$ terms which can be very small for thin shells.  Floating point
      errors are magnified when $m_p$ is multiplied by $h^2$.  In addition,
      it is not desirable to interpolate the plate thickness to the grid.
    \item  {\bf Compute heat and momentum exchange due to contact.}\\
      In this step, any heat and momentum exchange between bodies inside the
      computational domain is performed through the grid.  Details of 
      contact algorithms used my the material point method can be found in
      references~\cite{Sulsky94, Bard00, Bard01}.
    \item  {\bf Compute the stress tensor.}\\
      The stress tensor computation follows the procedure for hyperelastic
      materials cited in reference~\cite{Simo98}.  However, some extra steps
      are required for shell materials.  The stress update is performed using 
      a forward Euler explicit time stepping procedure.  The velocity 
      gradient $\boldsymbol{\nabla}\mathbf{w}$ at a material point is required 
      for the stress update.  This quantity is determined using 
      equation~(\ref{eq:velGrad}).  The velocity gradient of the center
      of mass of the shell ($\boldsymbol{\nabla}\mathbf{u}$) is computed from 
      the grid velocities using gradient weighting functions of the form
      \begin{equation}
        \boldsymbol{\nabla} S^{(1)}_{p,v} = \frac{1}{V_p}
          \int_{\Omega_p\cap\Omega} \chi_p(\mathbf{x})~
            \boldsymbol{\nabla} S_v(\mathbf{x})~d{\mathbf{x}}
      \end{equation}
      so that
      \begin{equation}      
        \boldsymbol{\nabla}\mathbf{u}_p = \sum_v\mathbf{u}_v~
          \boldsymbol{\nabla} S^{(1)}_{p,v}~.
      \end{equation}      
      The gradient of the rotation rate ($\boldsymbol{\nabla}\mathbf{r}$) is
      also interpolated to the particles using the same procedure, i.e.,
      \begin{equation}      
        \boldsymbol{\nabla}\mathbf{r}_p = \sum_v\mathbf{r}_v~
          \boldsymbol{\nabla} S^{(1)}_{p,v}~.
      \end{equation}      
      The next step is to calculate the in-surface gradients 
      $\boldsymbol{\nabla}^{(s)}\mathbf{u}_p$ and 
      $\boldsymbol{\nabla}^{(s)}\mathbf{r}_p$.  These are calculated as
      \begin{eqnarray}
        \boldsymbol{\nabla}^{(s)}\mathbf{u}_p &=
          \boldsymbol{\nabla}\mathbf{u}_p \bullet 
          \left(\mathbf{I} - \mathbf{n}^n_p\otimes\mathbf{n}^n_p\right) \\
        \boldsymbol{\nabla}^{(s)}\mathbf{r}_p &=
          \boldsymbol{\nabla}\mathbf{r}_p \bullet 
          \left(\mathbf{I} - \mathbf{n}^n_p\otimes\mathbf{n}^n_p\right)
      \end{eqnarray}
      The superscript $n$ represents the values at the end of the $n$-th time 
      step.  The shell is now divided into a number of layers with different 
      values of $z$ (these can be considered to be equivalent to Gauss points to
      be used in the integration over $z$).  The number of layers depends
      on the requirements of the problem.  Three layers are used to obtain the 
      results that follow.  
      The velocity gradient $\boldsymbol{\nabla}\mathbf{w}_p$ is calculated for
      each of the layers using equation~(\ref{eq:velGrad}).  For a shell with 
      three layers (top, center and bottom), the velocity gradients are
      given by
      \begin{align}
        \boldsymbol{\nabla}\mathbf{w}_p^{\text{top}} & = 
          \left[\boldsymbol{\nabla}^{(s)}\mathbf{u}_p+
          h^+~\boldsymbol{\nabla}^{(s)}\mathbf{r}_p\right] + 
          \mathbf{r}_p^n\otimes\mathbf{n}_p^n \\
        \boldsymbol{\nabla}\mathbf{w}_p^{\text{cen}} & = 
          \boldsymbol{\nabla}^{(s)}\mathbf{u}_p+
          \mathbf{r}_p^n\otimes\mathbf{n}_p^n \\
        \boldsymbol{\nabla}\mathbf{w}_p^{\text{bot}} & = 
          \left[\boldsymbol{\nabla}^{(s)}\mathbf{u}_p-
          h^-~\boldsymbol{\nabla}^{(s)}\mathbf{r}_p\right] + 
          \mathbf{r}_p^n\otimes\mathbf{n}_p^n 
      \end{align}
      The increment of deformation gradient ($\Delta\mathbf{F}$) in each layer 
      is computed using
      \begin{equation}
        \Delta\mathbf{F}_p = 
          \Delta t~\boldsymbol{\nabla}\mathbf{w}_p + \mathbf{I}
      \end{equation}
      The total deformation gradient ($\mathbf{F}$) in each layer is 
      updated using
      \begin{equation}
        \boldsymbol{\tilde{\mathbf{F}}}_p^{n+1} = 
          \Delta\mathbf{F}_p\bullet\mathbf{F}_p^{n}
      \end{equation}
      where $\boldsymbol{\tilde{\mathbf{F}}}_p^{n+1}$ is the intermediate 
      updated deformation gradient prior to application of the ``plane stress''
      condition.

      The stress in the shell is computed using a stored energy function ($W$)
      of the form
      \begin{equation}
        W = \frac{1}{2}K\left[\frac{1}{2}(J^2-1) - \ln J\right] +
            \frac{1}{2}G \left[\text{tr}(\boldsymbol{\Bar{\mathbf{b}}})-3\right]
      \end{equation}
      where $K$ is the bulk modulus, $G$ is the shear modulus, $J$ is
      the Jacobian ($J = \text{det}~\mathbf{F}$), 
      and $\boldsymbol{\Bar{\mathbf{b}}}$ is the 
      volume preserving part of the left Cauchy-Green strain tensor, defined as
      \begin{equation}
        \boldsymbol{\Bar{\mathbf{b}}} \colon= J^{-\frac{2}{3}} 
          \mathbf{F}\bullet\mathbf{F}^T
      \end{equation}
      The Cauchy stress then has the form
      \begin{equation}
        \boldsymbol{\sigma} = \frac{1}{2}K\left(J-\frac{1}{J}\right)\mathbf{I} +
           \frac{G}{J}\left[\boldsymbol{\Bar{\mathbf{b}}} - 
             \frac{1}{3}\text{tr}(\boldsymbol{\Bar{\mathbf{b}}})\right]~.
           \label{eq:elastic}
      \end{equation}

      The ``plane stress'' condition in the thickness direction of the shell
      is applied at this stage using an iterative Newton method.  To apply
      this condition, the deformation gradient tensor has to be rotated such 
      that its $(33)$ component is aligned with the $(zz)$ direction of the 
      shell.  The rotation tensor is the one required to rotate
      the vector $\mathbf{e}_3 \equiv (0,0,1)$ to the direction 
      $\mathbf{n}_p^n$ about the vector $\mathbf{e}_3 \times \mathbf{n}_p^n$.
      If $\theta$ is the angle of rotation and $\mathbf{a}$ is the unit vector
      along axis of rotation, the rotation tensor is given by (using the 
      derivative of the Euler-Rodrigues formula)
      \begin{equation}
        \mathbf{R} = \cos\theta\left(\mathbf{I}-
           \mathbf{a}\otimes\mathbf{a}\right) + \mathbf{a}\otimes\mathbf{a} -
           \sin\theta~\mathbf{A}
         \label{eq:rotation}
      \end{equation}
      where 
      \begin{equation}
        \mathbf{A} = \begin{bmatrix} 0 & -a_3 & a_2 \\ a_3 & 0 & -a_1 \\
                       -a_2 & a_1 & 0 \end{bmatrix}~.
      \end{equation}
      The rotated deformation gradient in each layer is given by
      \begin{equation}
        \mathbf{F}_p^{\text{rot}} = \mathbf{R}\bullet
          \boldsymbol{\tilde{\mathbf{F}}}_p^{n+1}
          \bullet\mathbf{R}^T~.
      \end{equation}
      The updated stress ($\boldsymbol{\sigma}_p^{\text{rot}}$) is calculated 
      in this rotated coordinate system using
      equation~(\ref{eq:elastic}).  Thus, 
      \begin{equation}
        \boldsymbol{\sigma}_p^{\text{rot}} = 
           \frac{1}{2}K\left(J_p^{\text{rot}}-\frac{1}{J_p^{\text{rot}}}\right)
           \mathbf{I} +
           \frac{G}{J_p^{\text{rot}}}\left[
             \boldsymbol{\Bar{\mathbf{b}}}_p^{\text{rot}} - 
             \frac{1}{3}\text{tr}(
             \boldsymbol{\Bar{\mathbf{b}}}_p^{\text{rot}})\right]~.
      \end{equation}
      An iterative Newton method is used to
      determine the deformation gradient component $F_{33}$ for
      which the stress component $\sigma_{33}$ is zero.  The 
      ``plane stress'' deformation gradient is denoted 
      $\boldsymbol{\overset{\circ}{\mathbf{F}}}$ and the stress is denoted
      $\boldsymbol{\overset{\circ}{\sigma}}$.  

      At this stage, the updated
      thickness of the shell at a material point is calculated from the 
      relations
      \begin{align}
        h^+_{n+1} &= h^+_0 \int^1_0 \overset{\circ}{F}_{zz}(+z)~dz \\
        h^-_{n+1} &= h^-_0 \int^1_0 \overset{\circ}{F}_{zz}(-z)~dz 
      \end{align}
      where $h^+_0$ and $h^-_0$ are the initial values, and $h^+_{n+1}$ and 
      $h^-_{n+1}$ are the updated values, of $h^+$ and $h^-$, respectively.

      In the next step, the deformation gradient and stress values
      for all the layers at each material point are rotated back to the
      original coordinate system.  The updated Cauchy stress and deformation
      gradient are
      \begin{align}
        \mathbf{F}_p^{n+1} & = \mathbf{R}^T\bullet
           \boldsymbol{\overset{\circ}{\mathbf{F}}} \bullet\mathbf{R} \\
        \boldsymbol{\sigma}_p^{n+1} & = \mathbf{R}^T\bullet
           \boldsymbol{\overset{\circ}{\sigma}} \bullet\mathbf{R}~.
      \end{align}
      The deformed volume of the shell is approximated using the Jacobian of 
      the deformation gradient at the center of mass of the shell 
      \begin{equation}
         V_p^{n+1} = V_p^0~J_p^{n+1}~.
      \end{equation}
    \item  {\bf Compute the internal force and moment.}\\
      The internal force for general materials is computed at the grid
      using the relation
      \begin{equation}
        \mathbf{f}_v^{\text{int}} = \sum_p \left[\boldsymbol{\sigma}_p^{n+1}
          \bullet \boldsymbol{\nabla}S^{(1)}_{p,v}\right]~V_{p}^{n+1}
      \end{equation}
      For shell materials, this relation takes the form
      \begin{equation}
        \mathbf{f}_v^{\text{int}} = \sum_p \left[\left<
          \boldsymbol{\sigma}_p^{n+1}\right>
          \bullet \boldsymbol{\nabla}S^{(1)}_{p,v}\right]~V_{p}^{n+1}
      \end{equation}
      
      In addition to internal forces, the formulation for shell materials
      requires the computation of internal moments in order to solve for
      the rotational acceleration in the rotational inertia equation
      ~(\ref{eq:rotInertia}).  To obtain the discretized form of equation
      ~(\ref{eq:rotInertia}), the equation is integrated over the volume
      of the shell leading to~\cite{Lewis98}
      \begin{equation}
      -\sum_p\left[\left(\mathbf{M}_p\bullet\boldsymbol{\nabla}S^{(1)}_{p,v}
        \bullet\mathbf{I}^{(s)}\right)+
        \left(\mathbf{n}_p\bullet\left<\boldsymbol{\sigma}_p\right>\bullet
       \mathbf{I}^{(s)}\right)~S^{(0)}_{p,v}\right]~V_p
      = \left(\frac{1}{12}\sum_p S^{(0)}_{p,v}~m_p~h^2_p\right)
        \dot{\mathbf{r}}_v~.
      \end{equation}
      The average stress over the thickness of the shell is calculated
      using equation~(\ref{eq:avStress}) and the average moment is 
      calculated using equation~(\ref{eq:avMoment}).  The trapezoidal rule is
      used in both cases.  Thus,
      \begin{align}
        \left<\boldsymbol{\sigma}_p^{n+1}\right> &=
             \frac{1}{h_{n+1}}\int^{h^+_{n+1}}_{-h^-_{n+1}} 
             \boldsymbol{\sigma}^{n+1}_p(z)~dz \\
        \mathbf{M}_p^{n+1} &= \mathbf{I}^{(s)}\bullet
              \left[ \frac{1}{h_{n+1}}\int^{h^+_{n+1}}_{-h^-_{n+1}}
              \boldsymbol{\sigma}_p^{n+1}(z)~z~dz
              \right]\bullet \mathbf{I}^{(s)}
      \end{align}
      where
      \begin{equation} 
        \mathbf{I}^{(s)} = \mathbf{I} - \mathbf{n}_p^n\otimes\mathbf{n}_p^n
      \end{equation} 
      These are required in the balance of rotational inertia that is used
      to compute the updated rotation rate and the updated director vector.
      The internal moment for the shell material points can therefore the
      calculated using
      \begin{equation}
        \mathbf{m}_v^{\text{int}} = 
          \sum_p\left[\left(\mathbf{M}_p^{n+1}\bullet
            \boldsymbol{\nabla}S^{(1)}_{p,v}
            \bullet\mathbf{I}^{(s)}\right)+
        \left(\mathbf{n}_p^{n}\bullet\left<\boldsymbol{\sigma}_p^{n+1}
          \right>\bullet
       \mathbf{I}^{(s)}\right)~S^{(0)}_{p,v}\right]~V_p^{n+1}
       \label{eq:intMoment}
      \end{equation}
      In practice, only the first term of equation~(\ref{eq:intMoment}) is
      interpolated to the grid and back to the particles.  The equation of 
      motion for rotational inertia is solved on the particles.
    \item  {\bf Solve the equations of motion.}\\
      The equations of motion for linear momentum are solved on the grid
      so that the acceleration at the grid vertices can be determined.
      The relation that is used is
      \begin{equation}
        \mathbf{\dot{u}}_v = \frac{1}{m_v}\left(\mathbf{f}_v^{\text{ext}} -
                                 \mathbf{f}_v^{\text{int}}\right)
      \end{equation}
      where $\mathbf{f}^{\text{ext}}$ are external forces.  

      The angular momentum equations are solved on the particles after
      interpolating the term 
      \begin{equation}
        \mathbf{\tilde{m}}_v = 
          \sum_p\left(\mathbf{M}_p^{n+1}\bullet
            \boldsymbol{\nabla}S^{(1)}_{p,v}
            \bullet\mathbf{I}^{(s)}\right)
      \end{equation}
      back to the material points to get $\mathbf{\tilde{m}}_p$.  The 
      rotational acceleration is calculated using
      \begin{equation}
        \dot{\mathbf{r}}_p = \left(\frac{12~V_p}{m_p~h^2_p}\right) 
          \left[\mathbf{m}_p^{\text{ext}} - \mathbf{\tilde{m}}_p
          - \mathbf{n}_p\bullet\left<\boldsymbol{\sigma}_p\right>\bullet
          \mathbf{I}^{(s)}\right]
        \label{eq:rotAcc}
      \end{equation}
    \item  {\bf Integrate the acceleration.}\\
       The linear acceleration in integrated using a forward Euler 
       rule on the grid, giving the updated velocity on the grid as
       \begin{equation} 
         \mathbf{u}_v^{n+1} = \mathbf{u}_v^n + \Delta t~\mathbf{\dot{u}}_v 
       \end{equation} 
       For the rotational acceleration, the same procedure is followed
       at each material point to obtain an intermediate increment
       \begin{equation} 
         \Delta\mathbf{\tilde{r}}_p = \Delta t~\mathbf{\dot{r}}_p 
       \end{equation} 
       The factor $m_p~h^2_p$ in the denominator of the right hand side of
       equation~(\ref{eq:rotAcc}) makes the differential equation stiff.
       An accurate solution of the equation requires an implicit integration
       or extremely small time steps.   Instead, an implicit correction is 
       made to $\Delta\mathbf{\tilde{r}}_p$ by solving the 
       equation~\cite{Kashiwa02}
       \begin{equation}
          \left[\mathbf{I} + 
            \beta\left(\mathbf{I} - \mathbf{n}_p^n\otimes\mathbf{n}_p^n\right)
            \right]
            \Delta\overset{\circ}{\mathbf{r}}_p = 
            \Delta\mathbf{\tilde{r}}_p
       \end{equation}
       where $\Delta\overset{\circ}{\mathbf{r}}_p$ is the corrected value
       of $\Delta\mathbf{\tilde{r}}_p$ and
       \begin{equation}
         \beta = \frac{6~E}{V_p~m_p}\left(\frac{\Delta t}{h}\right)^2
       \end{equation}
       which uses the Young's modulus $E$ of the shell material.
       The intermediate rotation rate is updated using the corrected 
       increment.  Thus,
       \begin{equation} 
         \mathbf{\overset{\star}{r}}_p^{n+1} = \mathbf{r}_p^n + 
            \Delta\overset{\circ}{\mathbf{r}}_p~.
       \end{equation} 
    \item {\bf Update the shell director and rotate the rotation rate}
       At this stage, the shell director at each material point is updated.
       The incremental rotation tensor $\Delta\mathbf{R}$ is calculated using
       equation~(\ref{eq:rotation}) with
       rotation angle $\theta = |r|\Delta t$ and axis of rotation
       \begin{equation}
         \mathbf{a} = \frac{\mathbf{n}_p^n \times 
                        \mathbf{\overset{\star}{r}}_p^{n+1}}
                        {|\mathbf{n}_p^n \times 
                        \mathbf{\overset{\star}{r}}_p^{n+1}|}~.
       \end{equation}
       The updated director is
       \begin{equation}
         \mathbf{n}_p^{n+1} = \Delta\mathbf{R}\bullet \mathbf{n}_p^{n}~.
       \end{equation}
       In addition, the rate of rotation has to be rotated so that the 
       direction is perpendicular to the director using,
       \begin{equation}
         \mathbf{r}_p^{n+1} = \Delta\mathbf{R}\bullet
            \mathbf{\overset{\star}{r}}_p^{n+1}~.
       \end{equation}
    \item  {\bf Interpolate back to the material points and update the state
           variables.}\\
       In the final step, the state variables at the grid are interpolated
       back to the material points using relations of the form
       \begin{equation}
         \mathbf{u}_p^{n+1} = \sum_v \mathbf{u}_v^{n+1} S^{(1)}_{p,v}
       \end{equation}
  \end{enumerate}
  Steps 1 through 8 are repeated for the next time step.

  \section{Results}
  Three tests of the shell formulation have been performed on different
  shell geometries - a plane shell, a cylindrical shell, and a spherical
  shell.
  \begin{enumerate}
    \item {\bf Punched Plane Shell}\\
    This problem involves the indentation of a plane, circular shell into
    a rigid cylindrical die of radius 8 cm.  The shell is made of annealed 
    copper with the properties and dimensions shown in 
    Table~\ref{tab:planeShell}.
    \begin{table}[h]
      \caption{Circular plane shell properties and dimensions.}
      \label{tab:planeShell}
      \begin{center}
      \begin{tabular}{cccccc}
         \hline
         $\rho_0$ & K      & G     & Thickness & Radius & Velocity \\
       (kg/m$^3$) & (GPa)  & (GPa) & (cm)      & (cm)   & (m/s) \\
         \hline
         8930     & 136.35 & 45.45 & 0.3       & 8      & 100 \\
         \hline
      \end{tabular}
      \end{center}
    \end{table}

    Snapshots of the deformation of the shell are shown in 
    Figure~\ref{fig:planeShell}.
    \begin{figure}[p]
      \begin{center}
        \scalebox{0.30}{\includegraphics{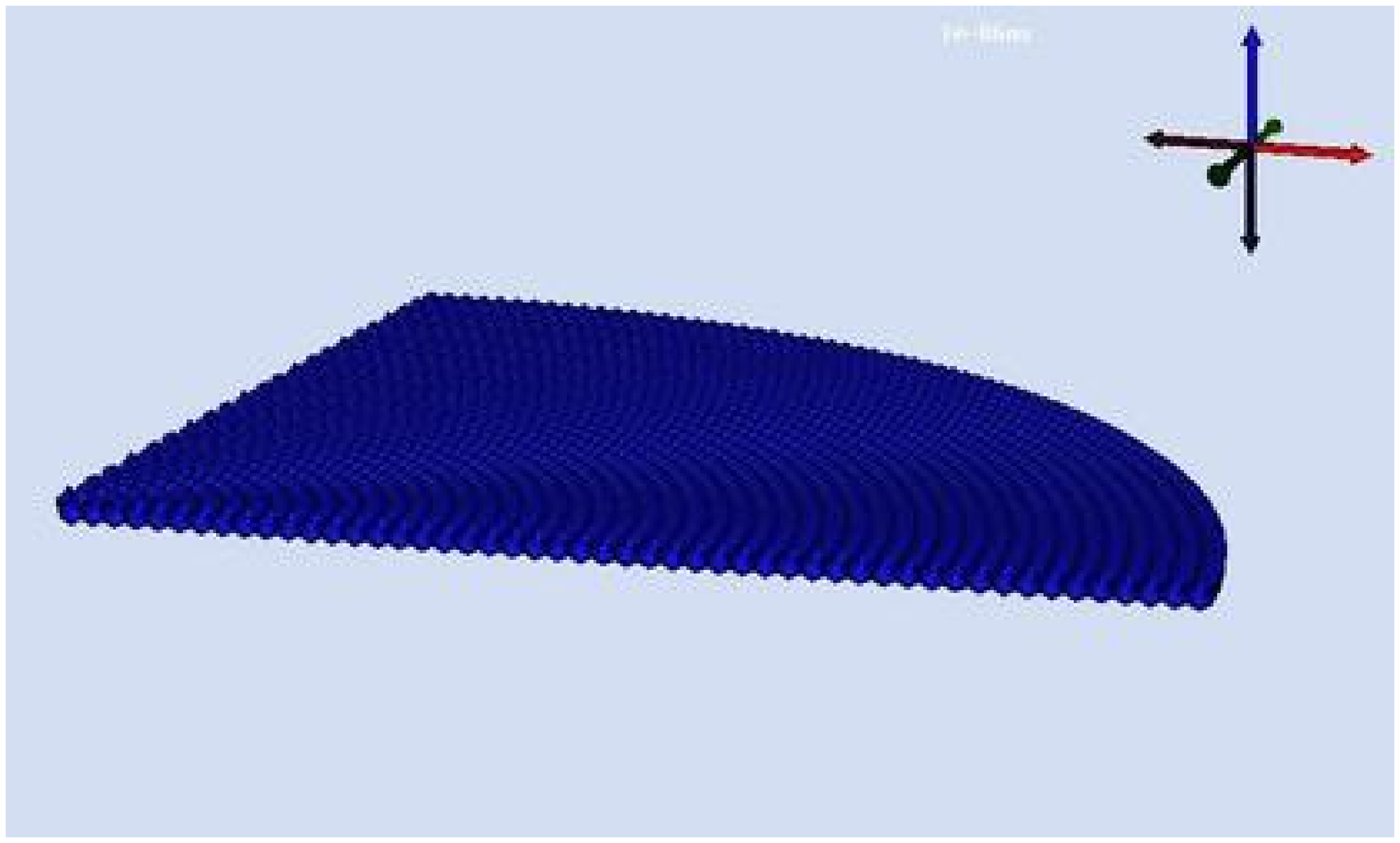}
                       \includegraphics{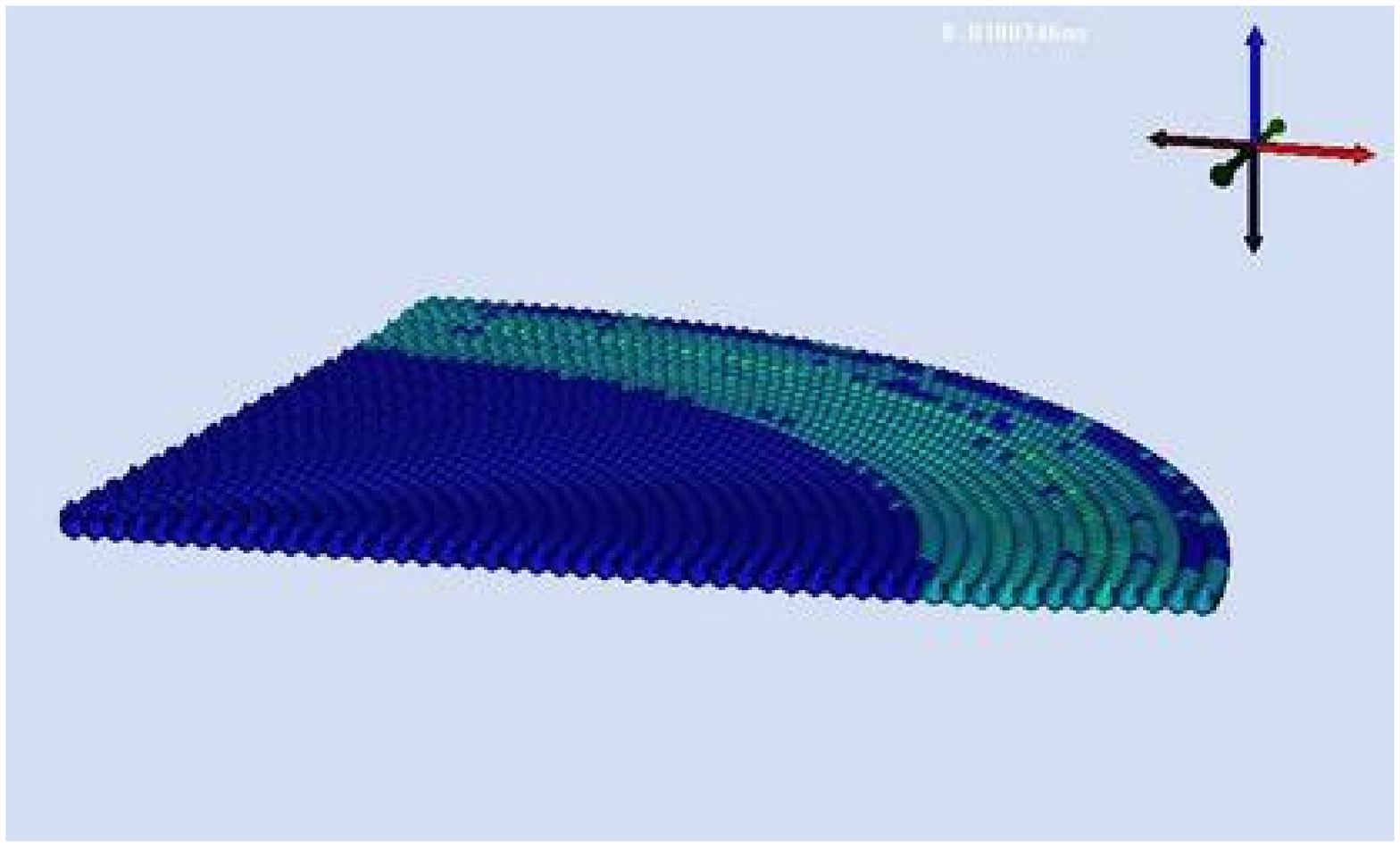}}
        \scalebox{0.30}{\includegraphics{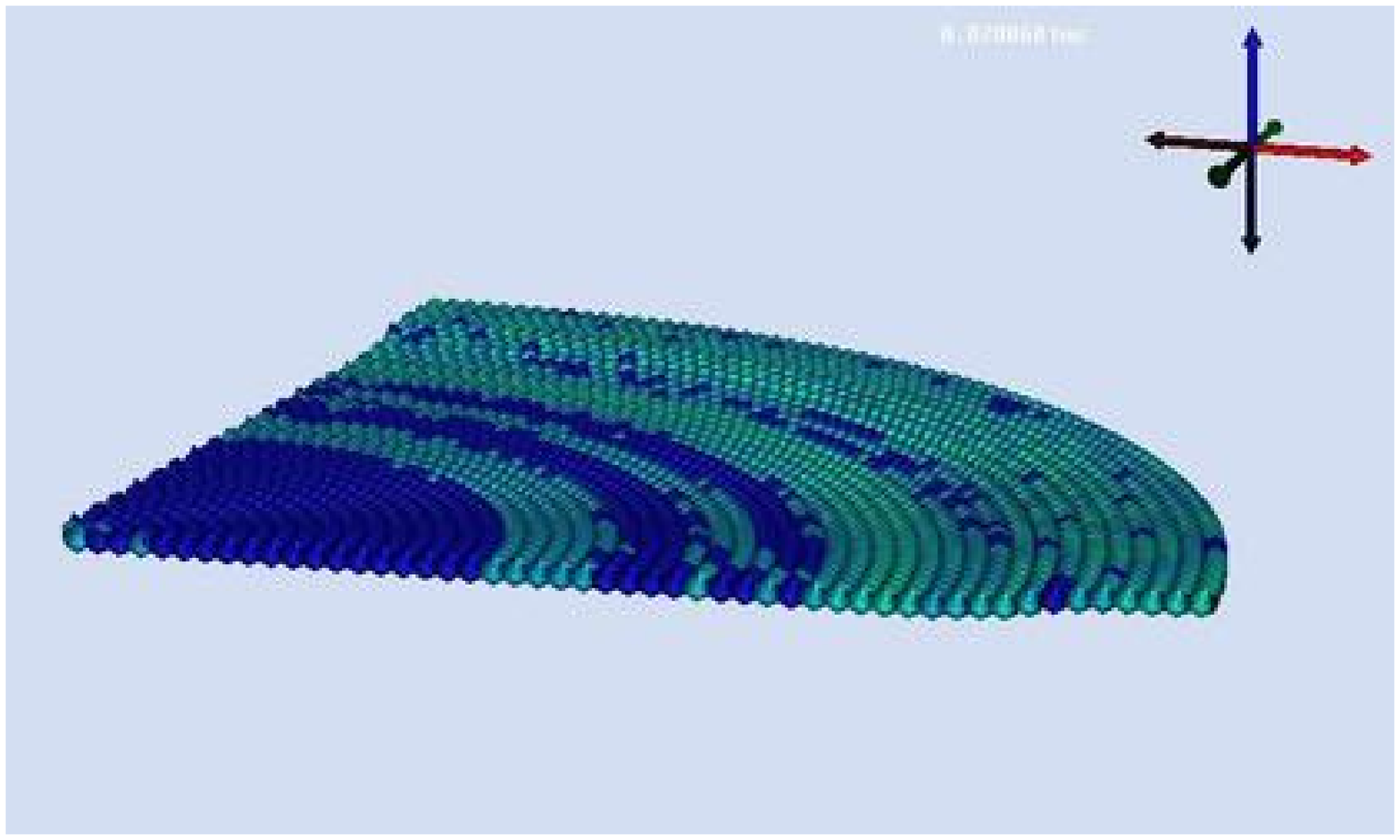}
                       \includegraphics{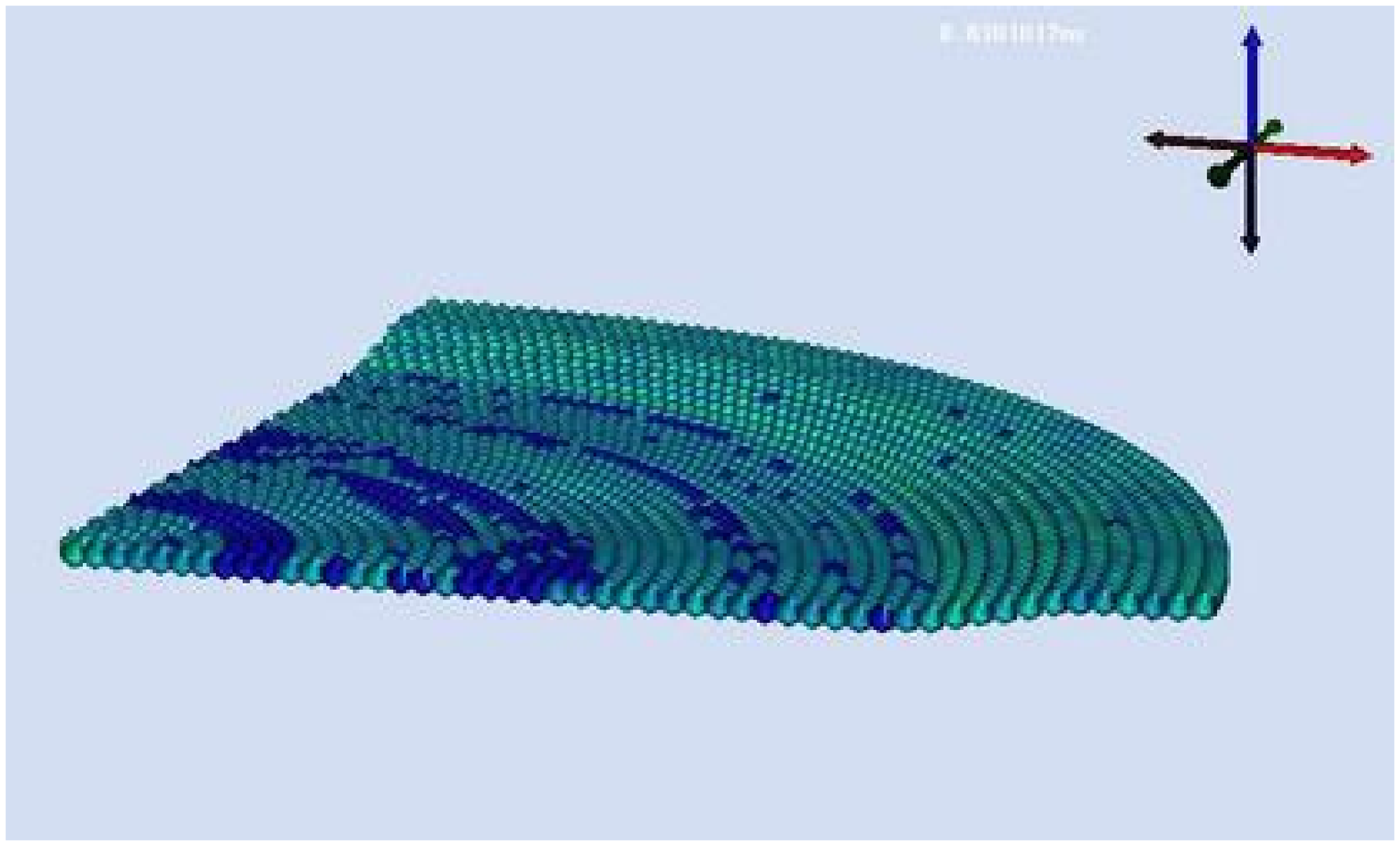}}
        \scalebox{0.30}{\includegraphics{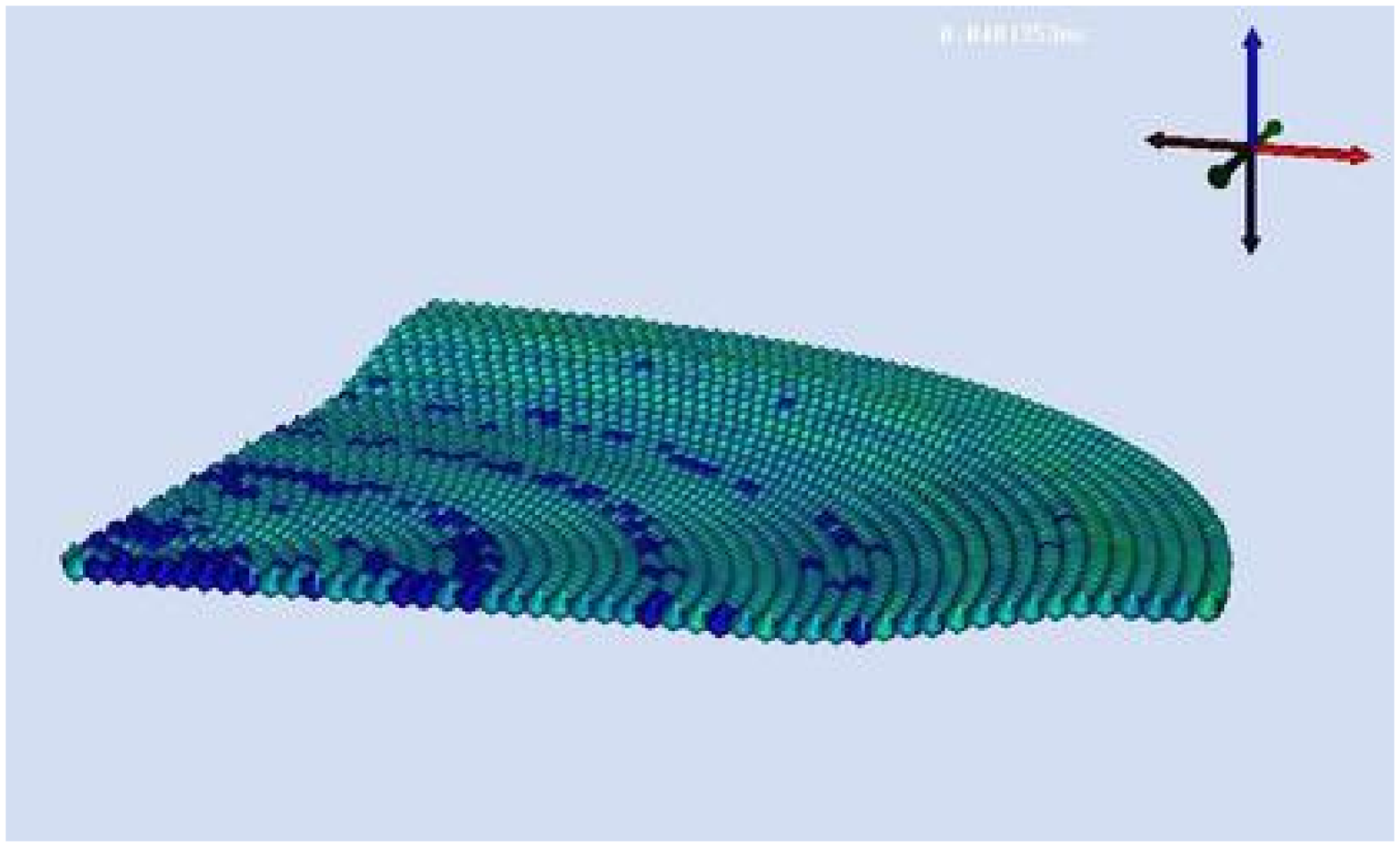}
                       \includegraphics{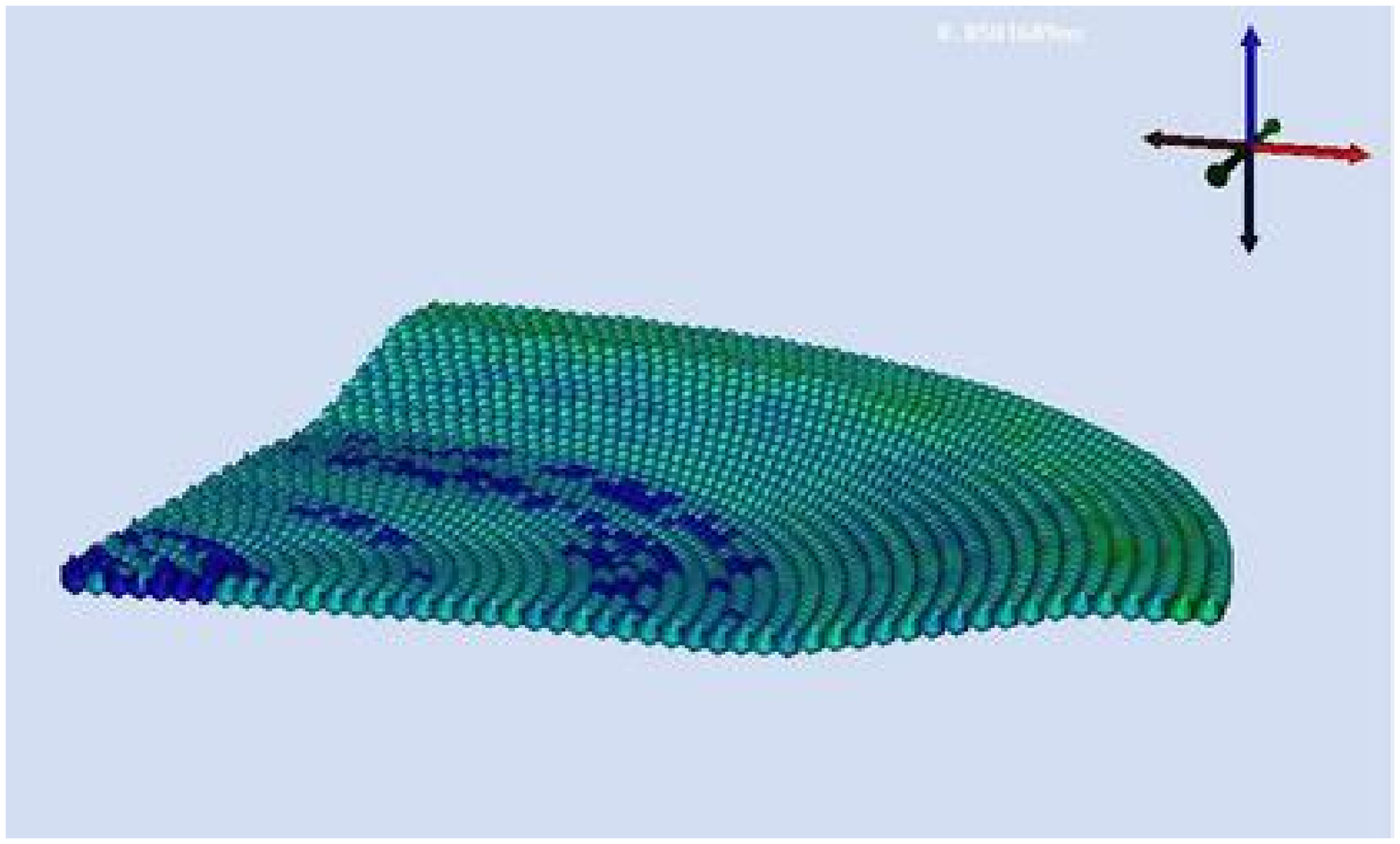}}
        \scalebox{0.30}{\includegraphics{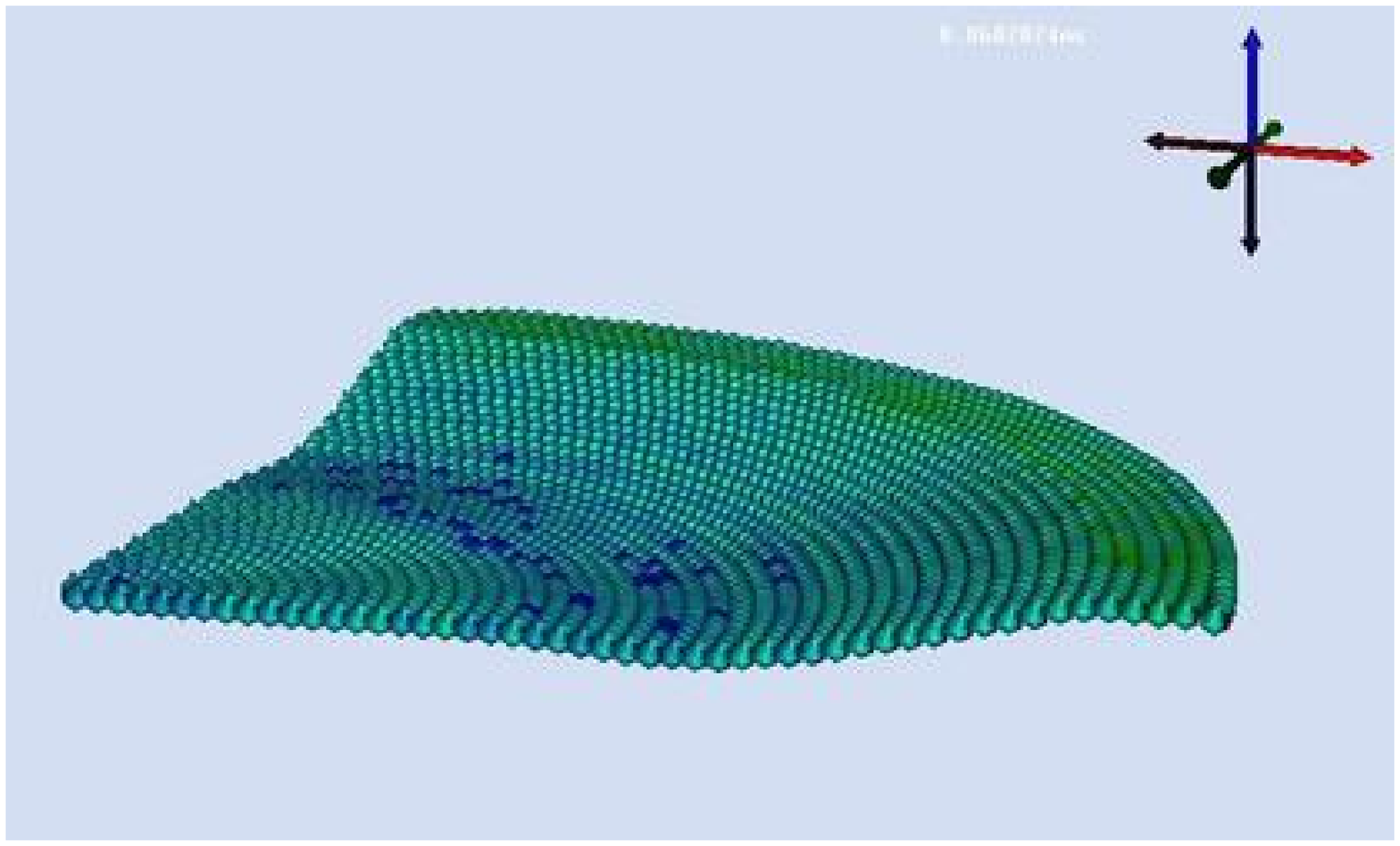}
                       \includegraphics{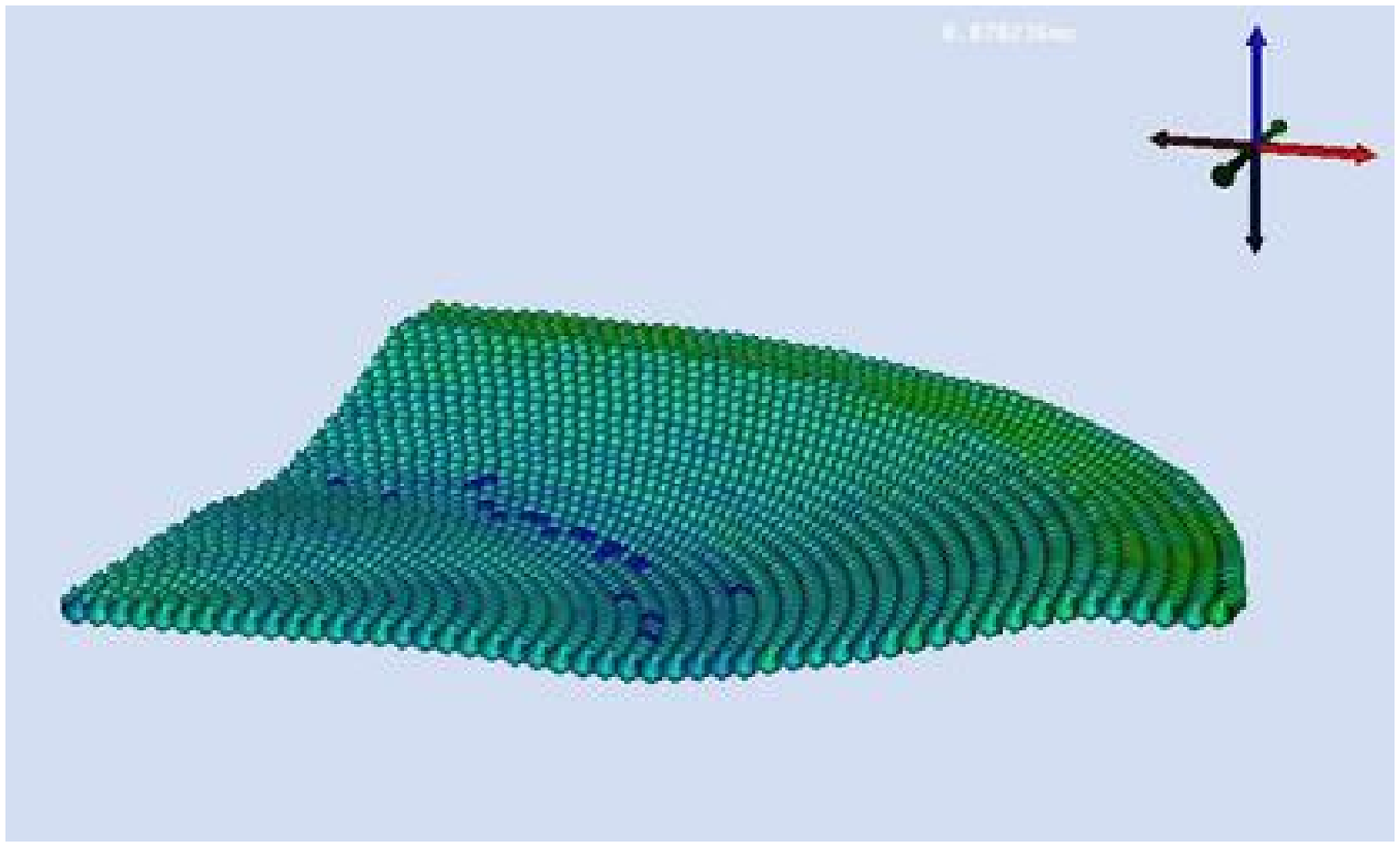}}
        \scalebox{0.30}{\includegraphics{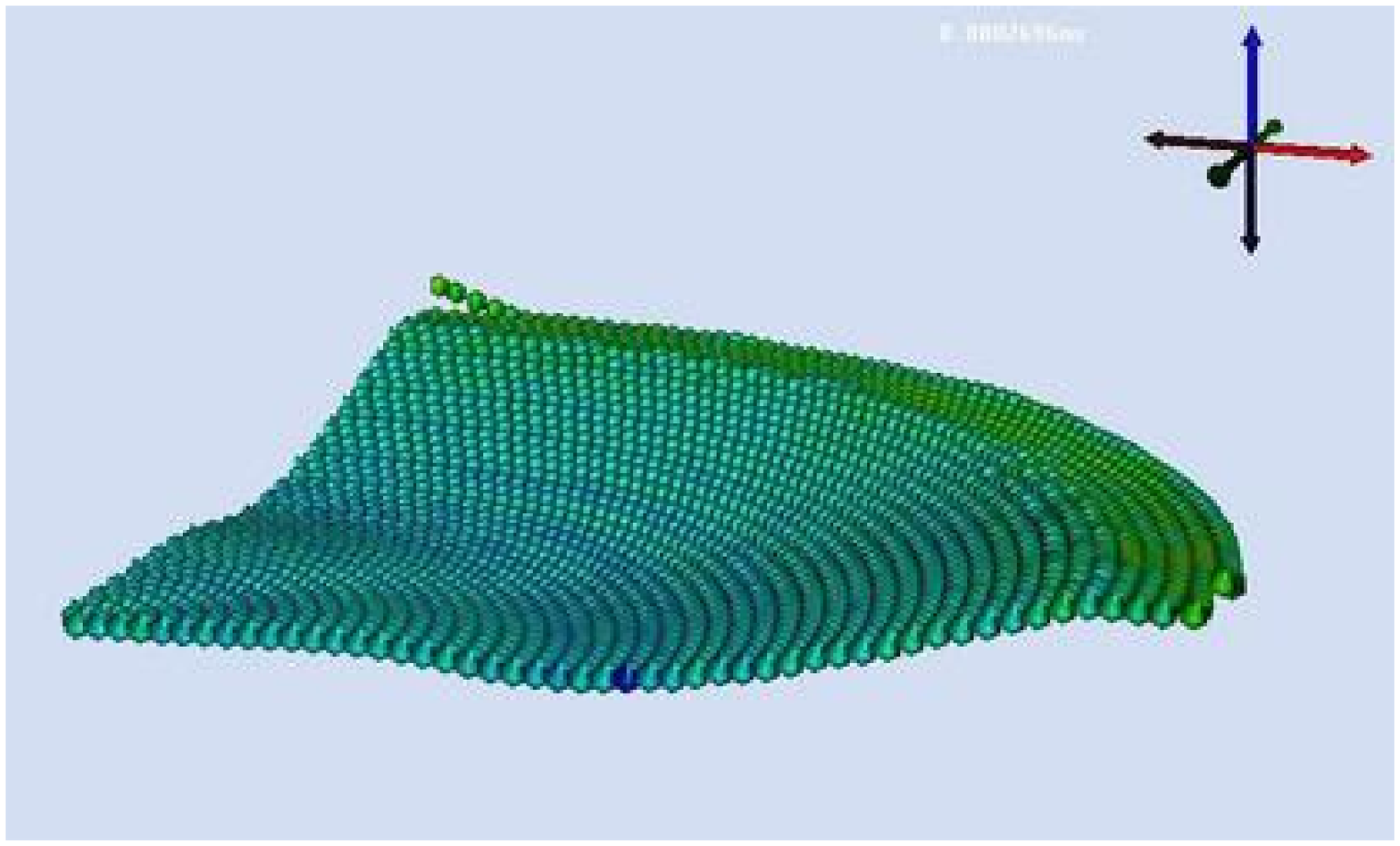}
                       \includegraphics{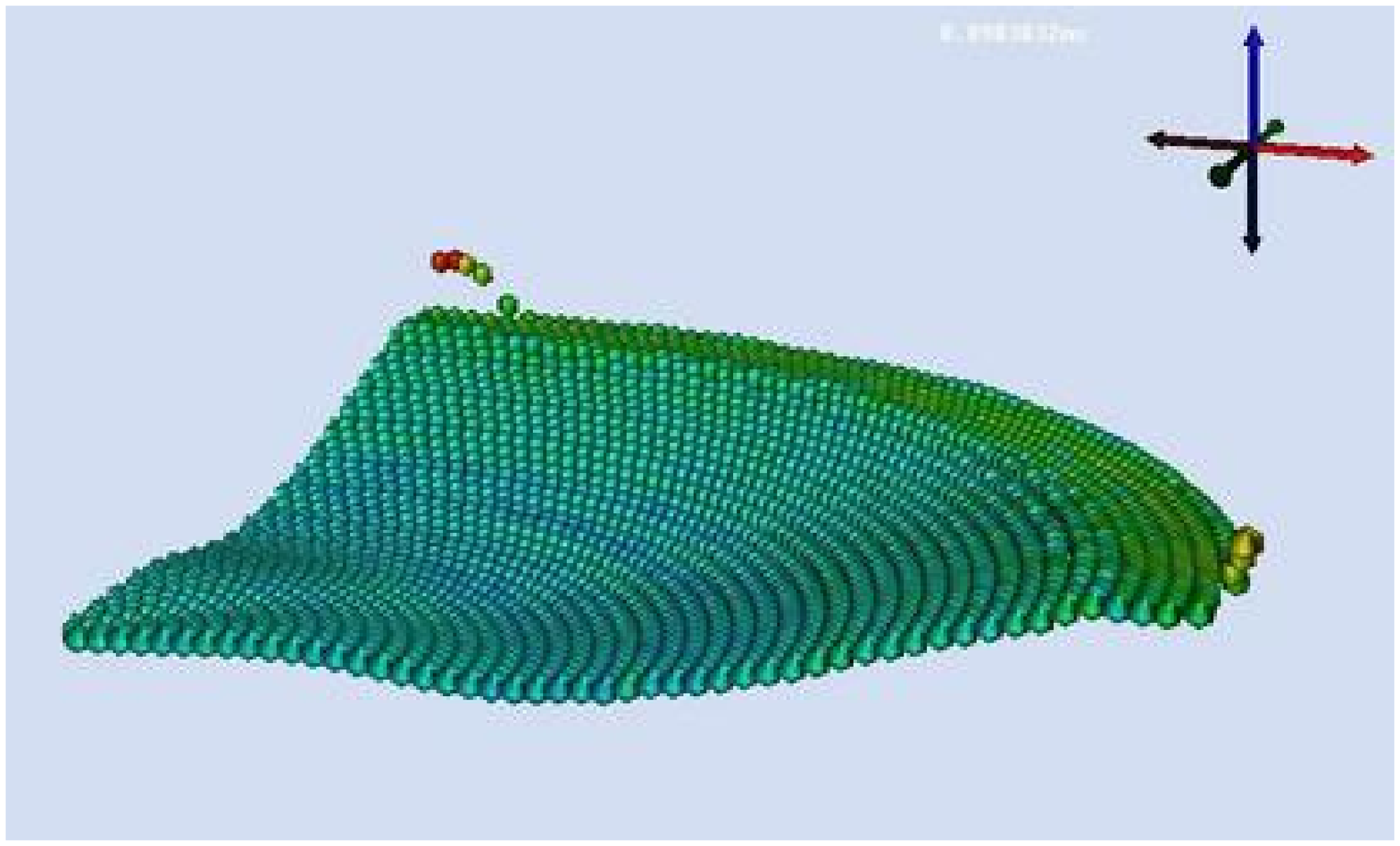}}
      \end{center}
      \caption{Deformation of punched circular plane shell.}
      \label{fig:planeShell}
    \end{figure}
    Substantial deformation of the shell occurs
    before particles at the edges tend to tear off.  The tearing off of 
    particles is due to the presence of large rotation rates ($\mathbf{r}$)
    which are due to the stiffness of the rotational acceleration equation
    ~(\ref{eq:rotAcc}).  The implicit correction does not appear to be
    adequate beyond a certain point and a fully implicit shell formulation
    may be required for accurate simulation of extremely large deformations.

    Particles in the figure have been colored using the equivalent stress at the
    center-of-mass of the shell.  The stress distribution in the shell is quite 
    uniform, though some artifacts in the form of rings appear.  An implicit 
    formulation has been shown to remove such artifacts in the stress 
    distribution in membranes~\cite{Guilkey02}.  Therefore, an implicit 
    formulation may be useful for the shell formulation.  Another possibility
    is that these artifacts may be due to membrane and shear locking, a known 
    phenomenon in finite element formulations of shells based on a continuum 
    approach~\cite{Belyt00, Libai98}.  Such locking effects can be reduced 
    using an addition hour glass control step~\cite{Belyt00} in the simulation. 
    \item {\bf Pinched Cylindrical Shell}\\
    The pinched cylindrical shell is one of the benchmark problems proposed
    by MacNeal and Harder~\cite{MacNeal85}.  The cylindrical shell that has
    been simulated in this work has dimensions similar to those used by
    Li et al.~\cite{LiShaofan00}.  The shell is pinched by contact with
    two small rigid solid cylinders placed diametrically opposite each other and
    located at the midpoint of the axis of the cylinder.  Each of the solid
    cylinders is 0.25 cm in radius, 0.5 cm in length, and moves toward the 
    center of the pinched shell in a radial direction at 10 ms$^{-1}$.  The 
    material of the shell is annealed copper (properties are shown in 
    Table~\ref{tab:planeShell}).  The cylindrical shell is 2.5 cm in radius,
    5.0 cm long, and 0.05 cm thick.
    
    Snapshots of the deformation of the pinched cylindrical shell are shown in 
    Figure~\ref{fig:cylShell}.
    \begin{figure}[p]
      \begin{center}
        \scalebox{0.30}{\includegraphics{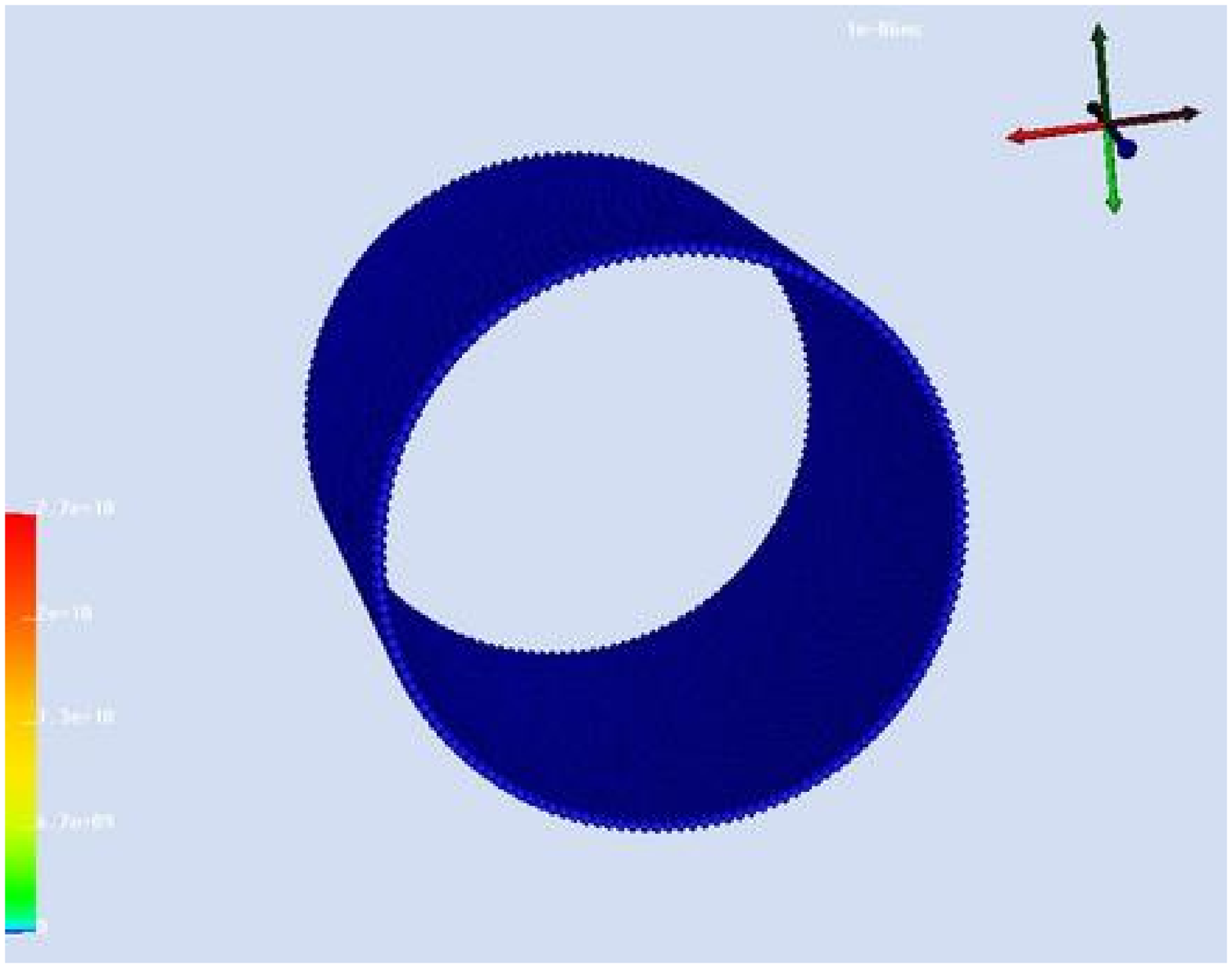}
                       \includegraphics{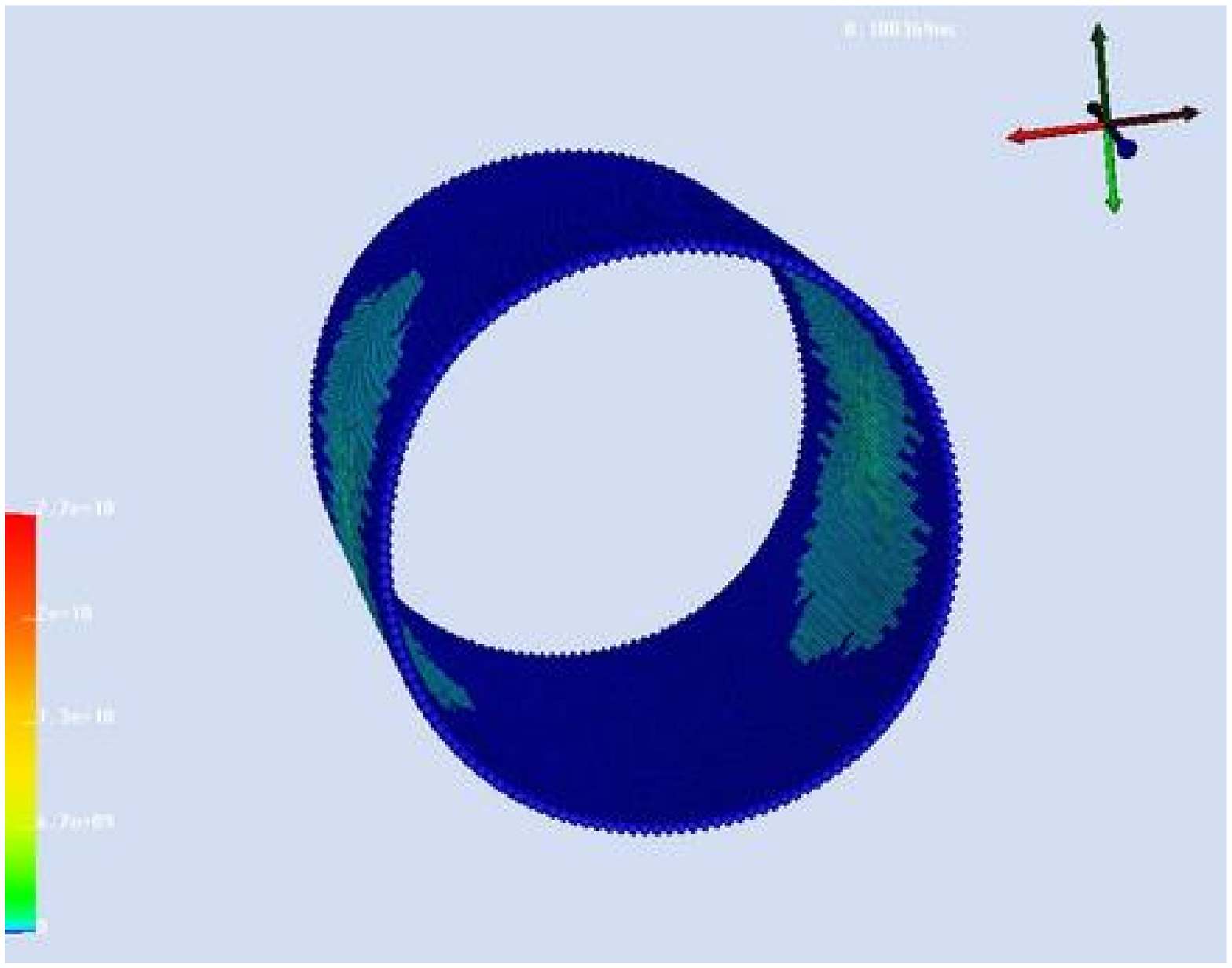}}
        \scalebox{0.30}{\includegraphics{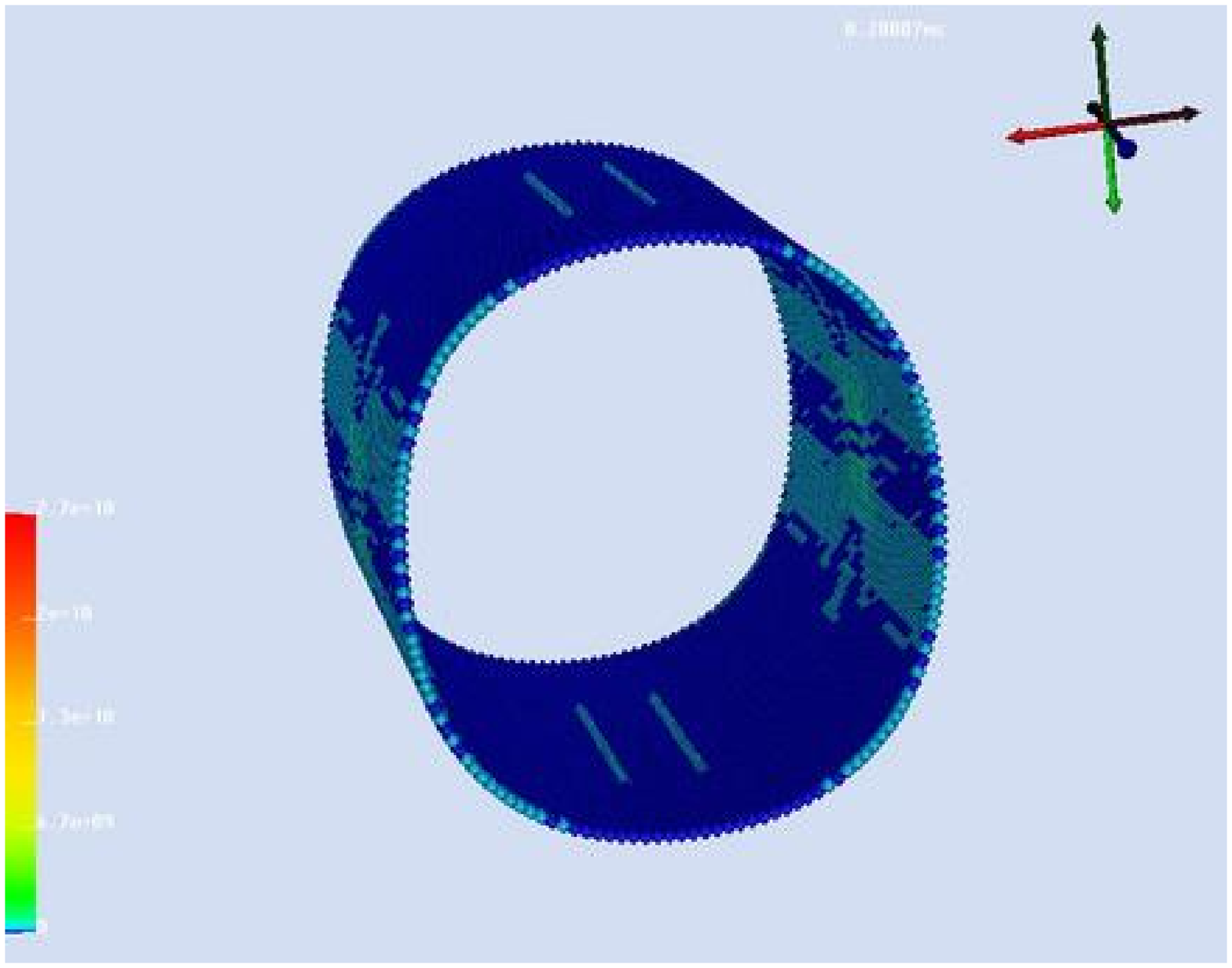}
                       \includegraphics{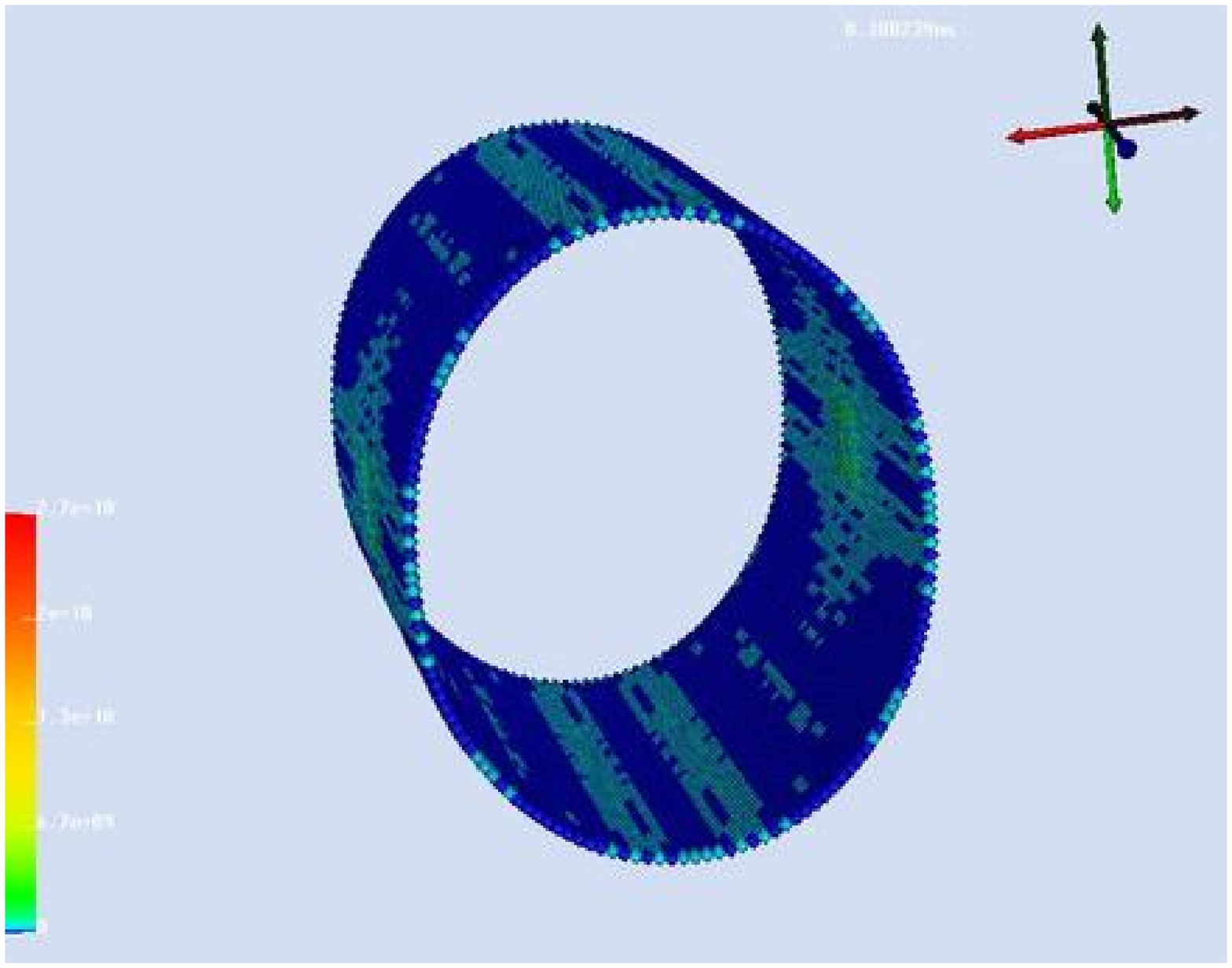}}
        \scalebox{0.30}{\includegraphics{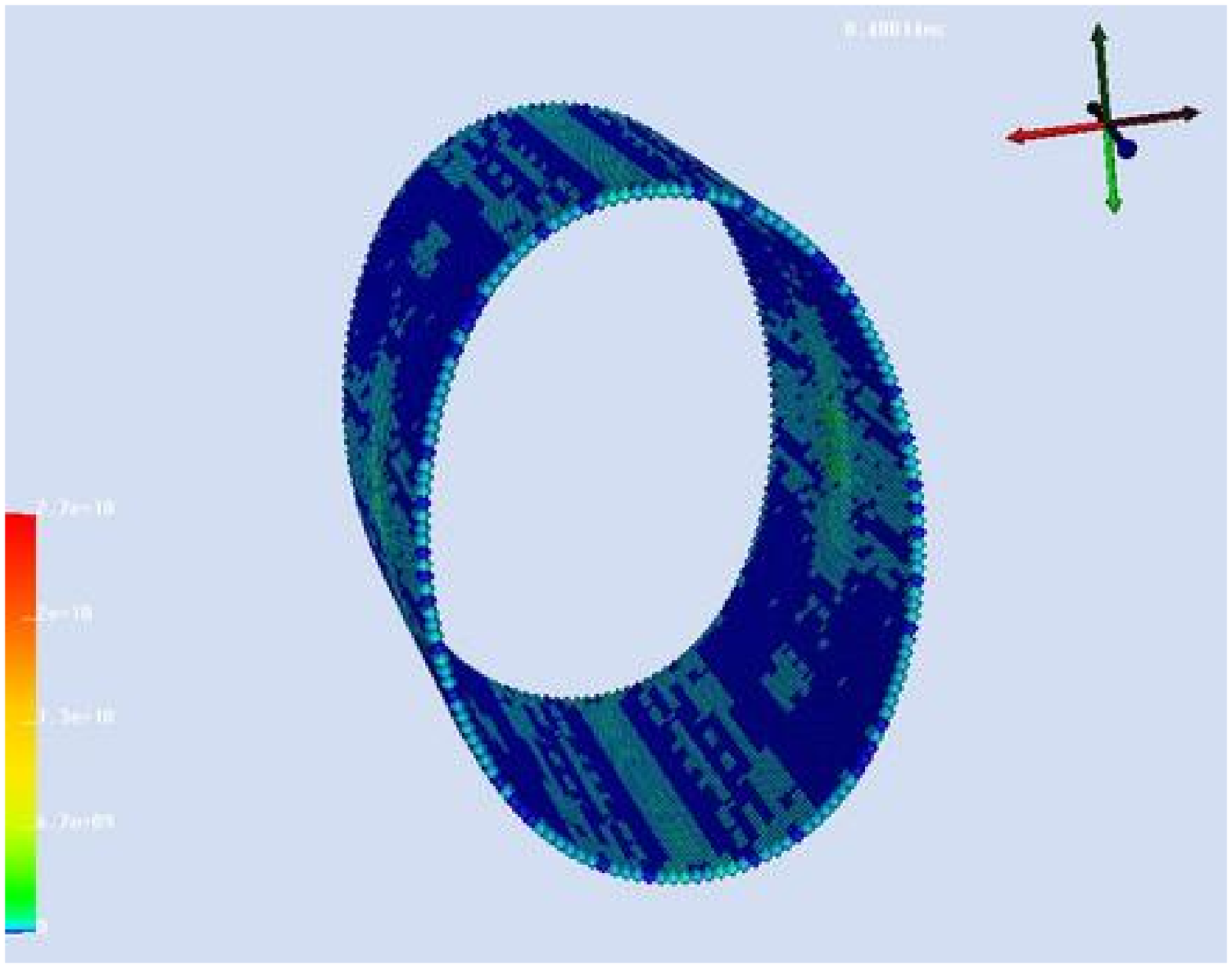}
                       \includegraphics{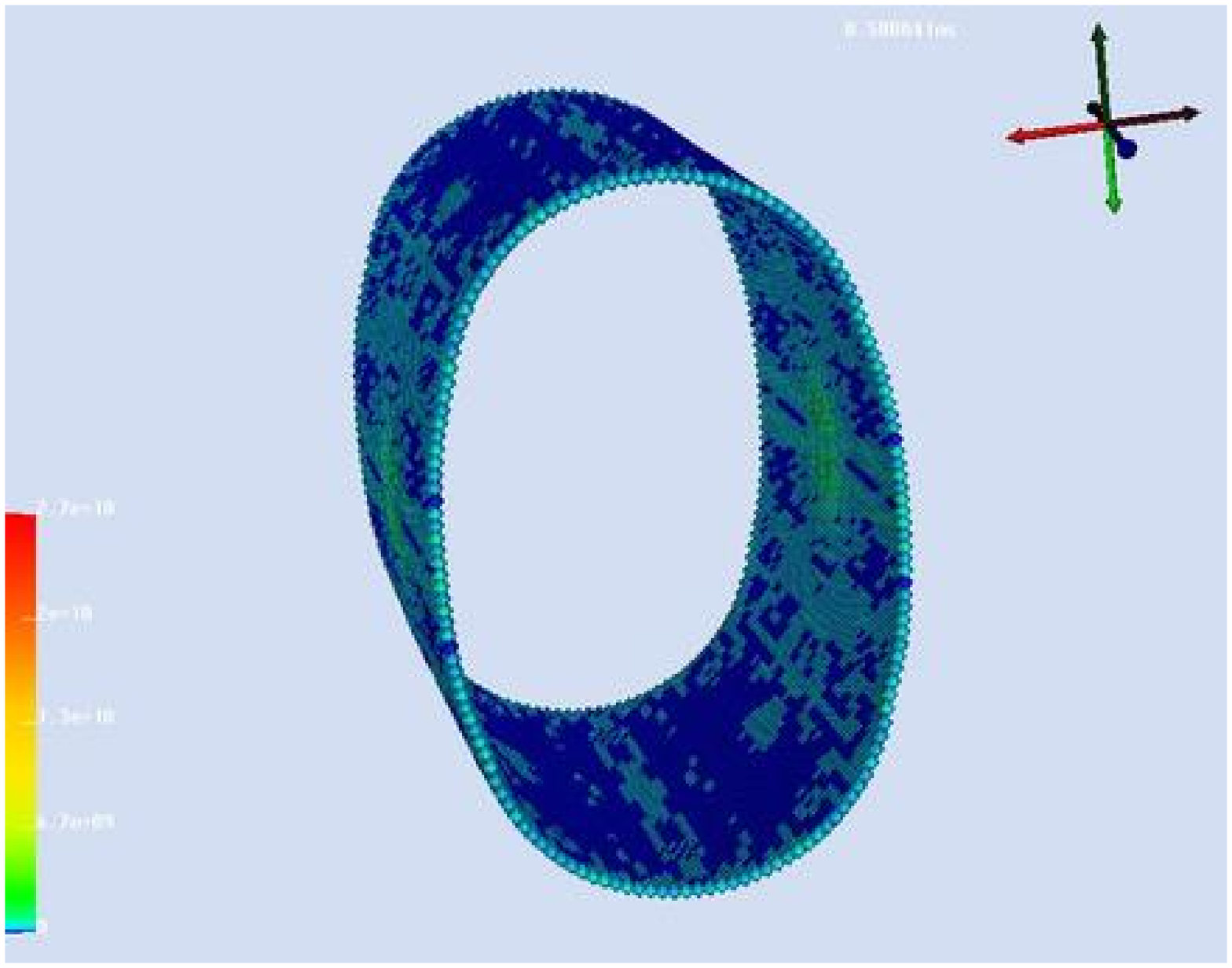}}
        \scalebox{0.30}{\includegraphics{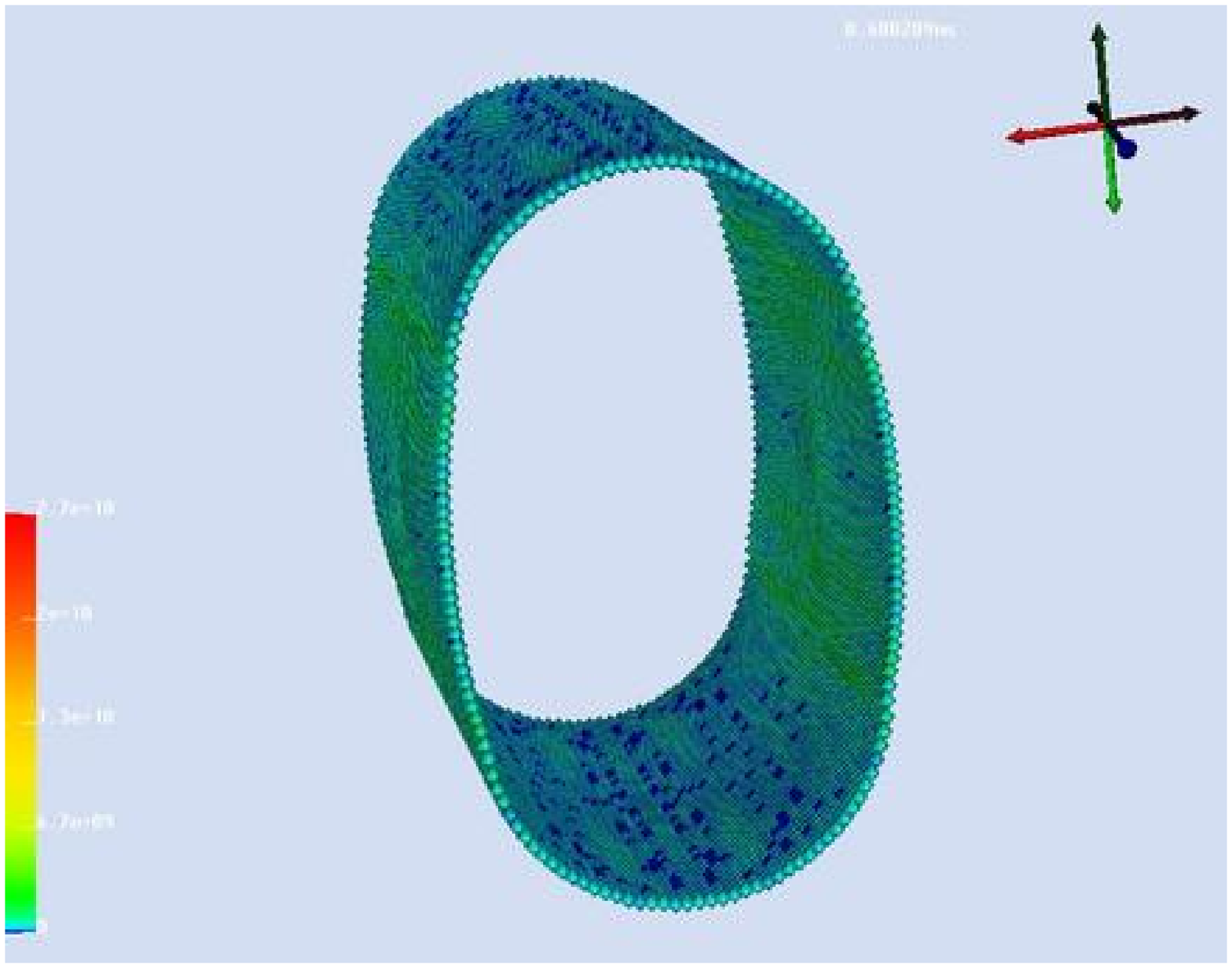}
                       \includegraphics{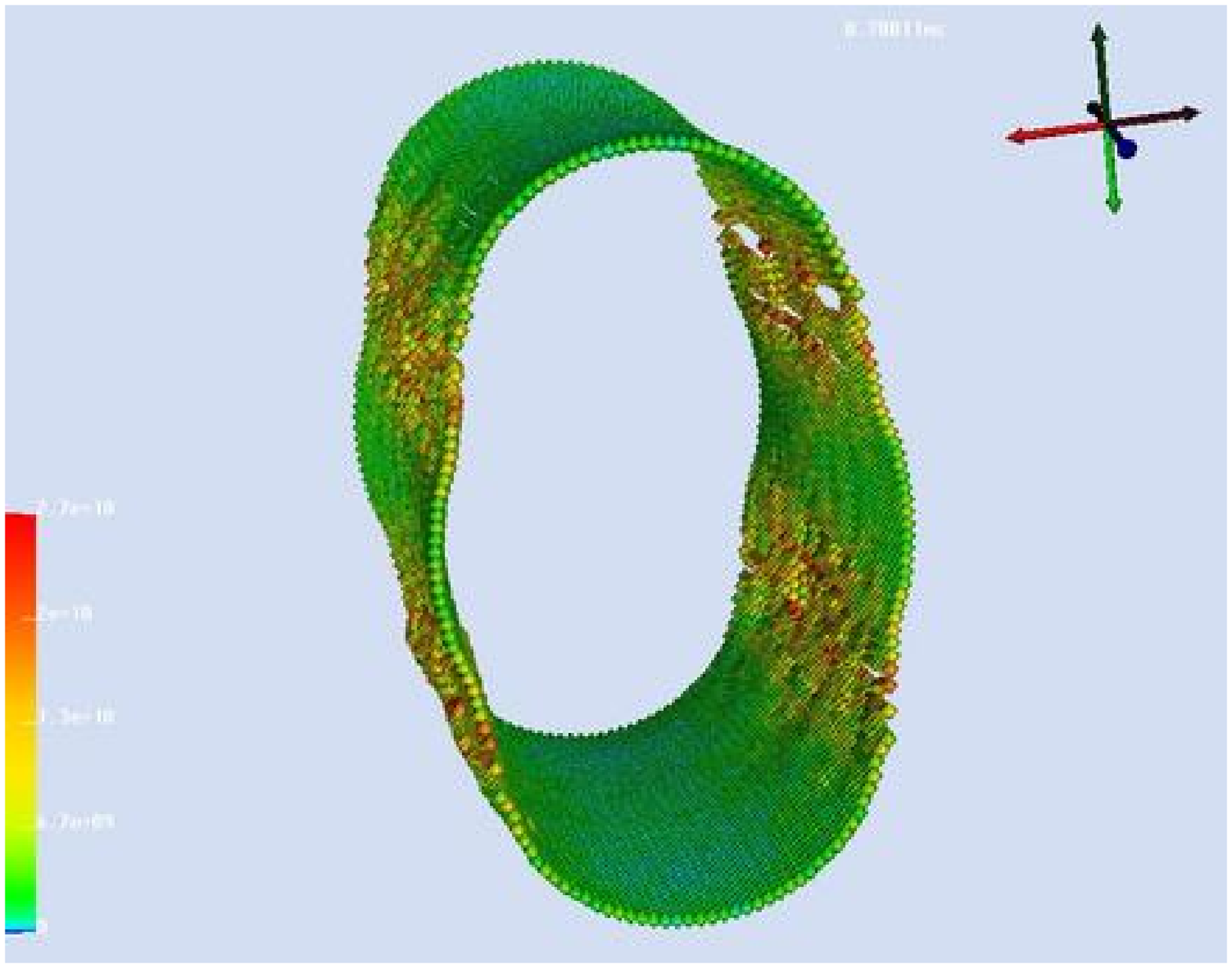}}
      \end{center}
      \caption{Deformation of pinched cylindrical shell.}
      \label{fig:cylShell}
    \end{figure}
    The deformation of the shell proceeds uniformly for 60 ms.  However,
    at this time the increments of rotation rate begin to increase rapidly at
    each time step, even though the velocity of the center-of-mass of the
    shell still remains stable.  This effect can be attributed to the stiffness
    of the rotational inertia equation.  The effect is that extremely
    large rotation rates are produced at 70 ms causing high velocities
    and eventual numerical fracture of the cylinder.  The problem may be
    solved using an implicit shell formulation.
    \item {\bf Inflating Spherical Shell}\\
    The inflating spherical shell problem is similar to that used to
    model lipid bilayers by Ayton et al.~\cite{Ayton02}.  The shell is
    made of a soft rubbery material with a density of 10 kg m$^{-3}$,
    a bulk modulus of 60 KPa and a shear modulus of 30 KPa.  The sphere
    has a radius of 0.5 m and is 1 cm thick.  The spherical shell is
    pressurized by an initial internal pressure of 10 KPa.  The pressure 
    increases in proportion to the internal surface area as the sphere inflates.

    The deformation of the shell with time is shown in 
    Figure~\ref{fig:sphShell}.
    \begin{figure}[p]
      \begin{center}
        \scalebox{0.30}{\includegraphics{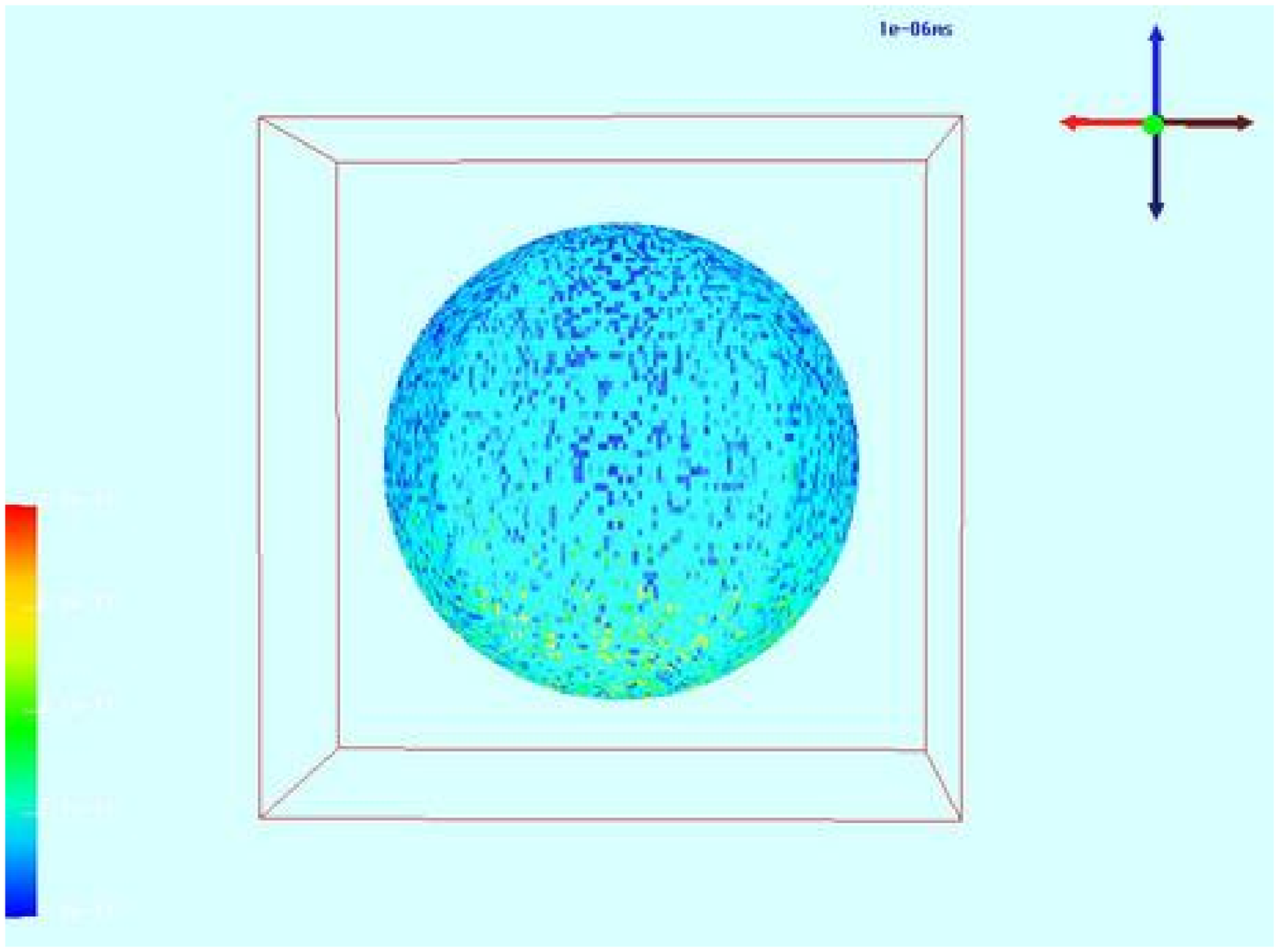}
                       \includegraphics{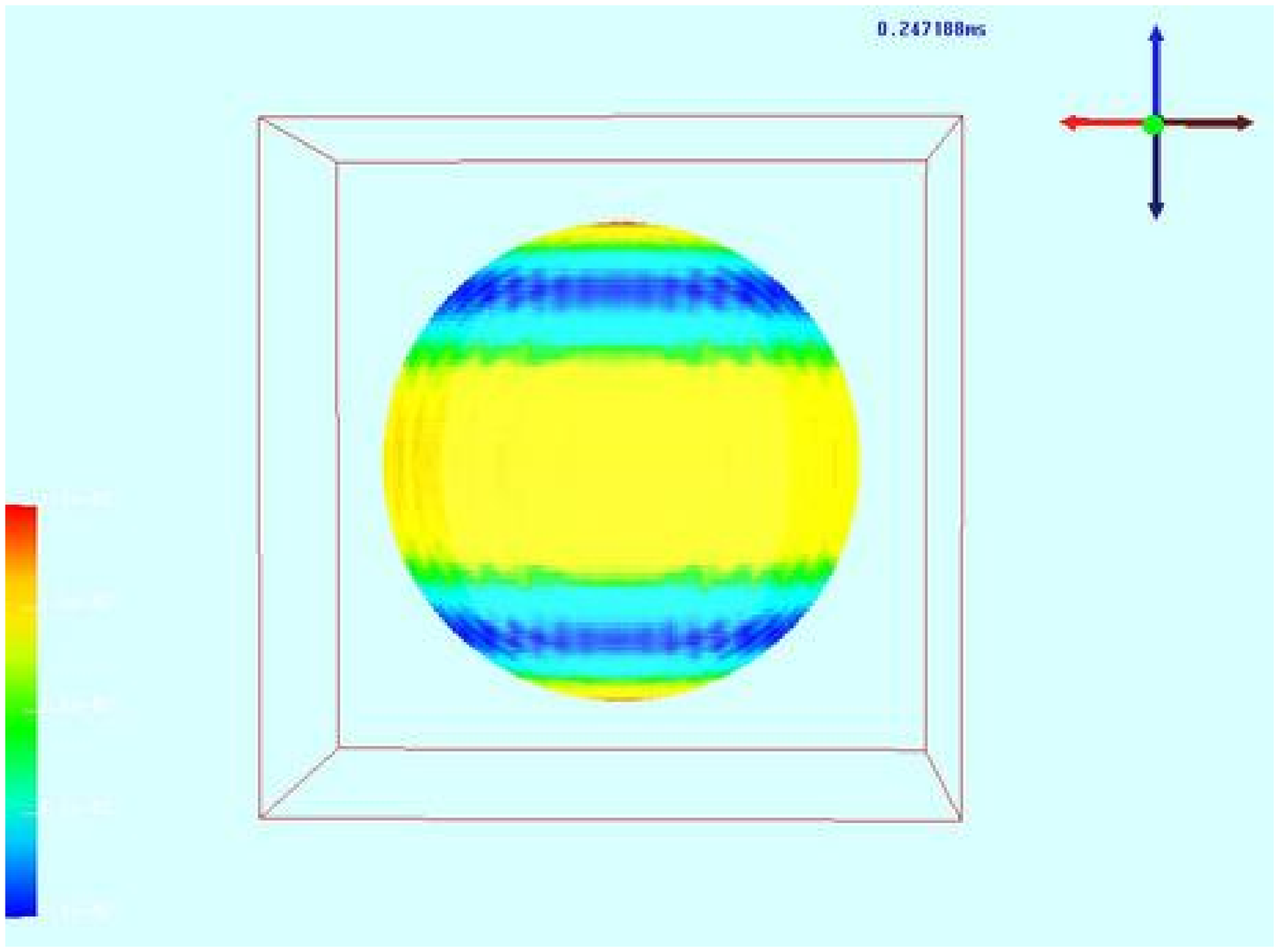}}
        \scalebox{0.30}{\includegraphics{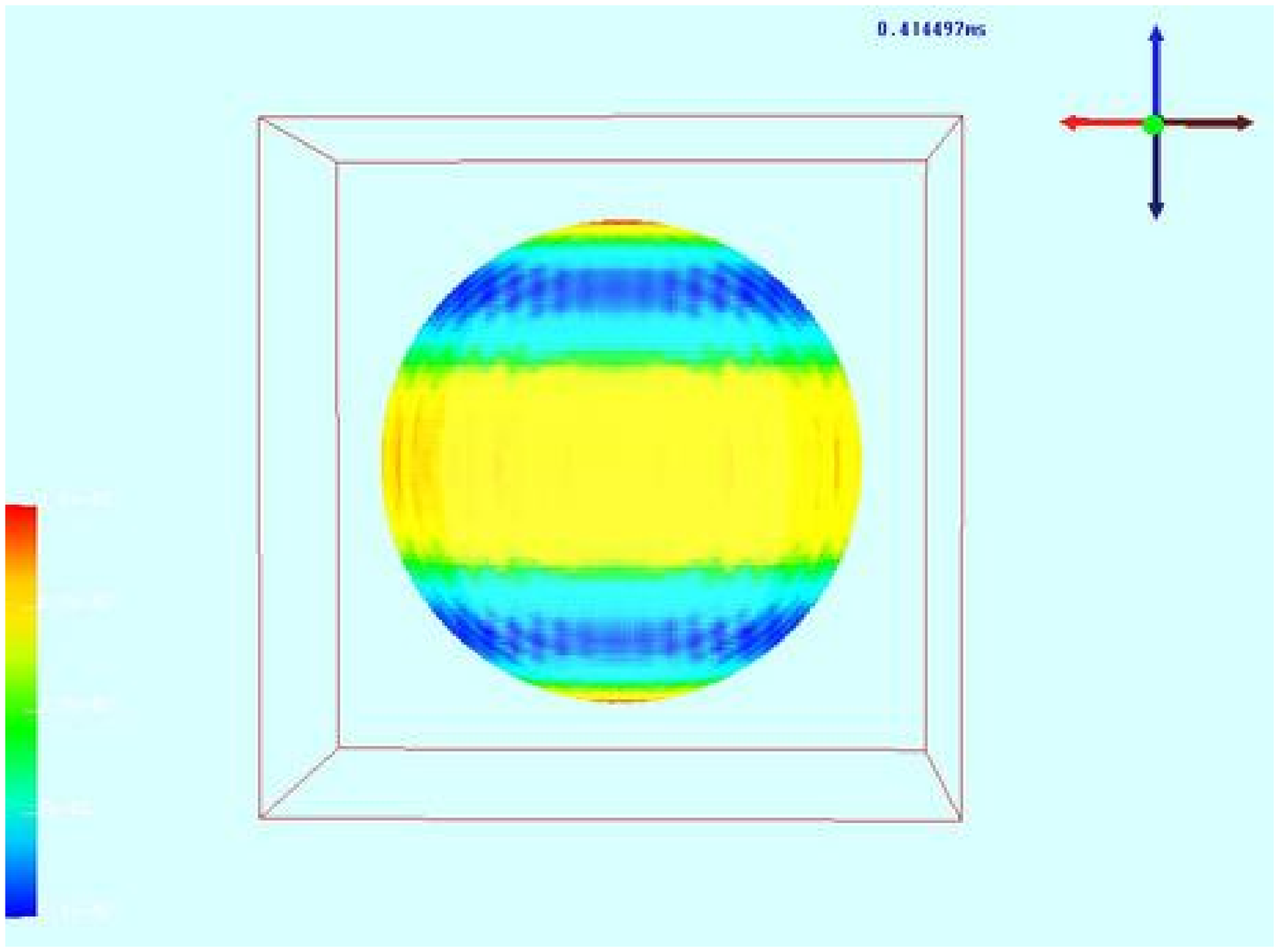}
                       \includegraphics{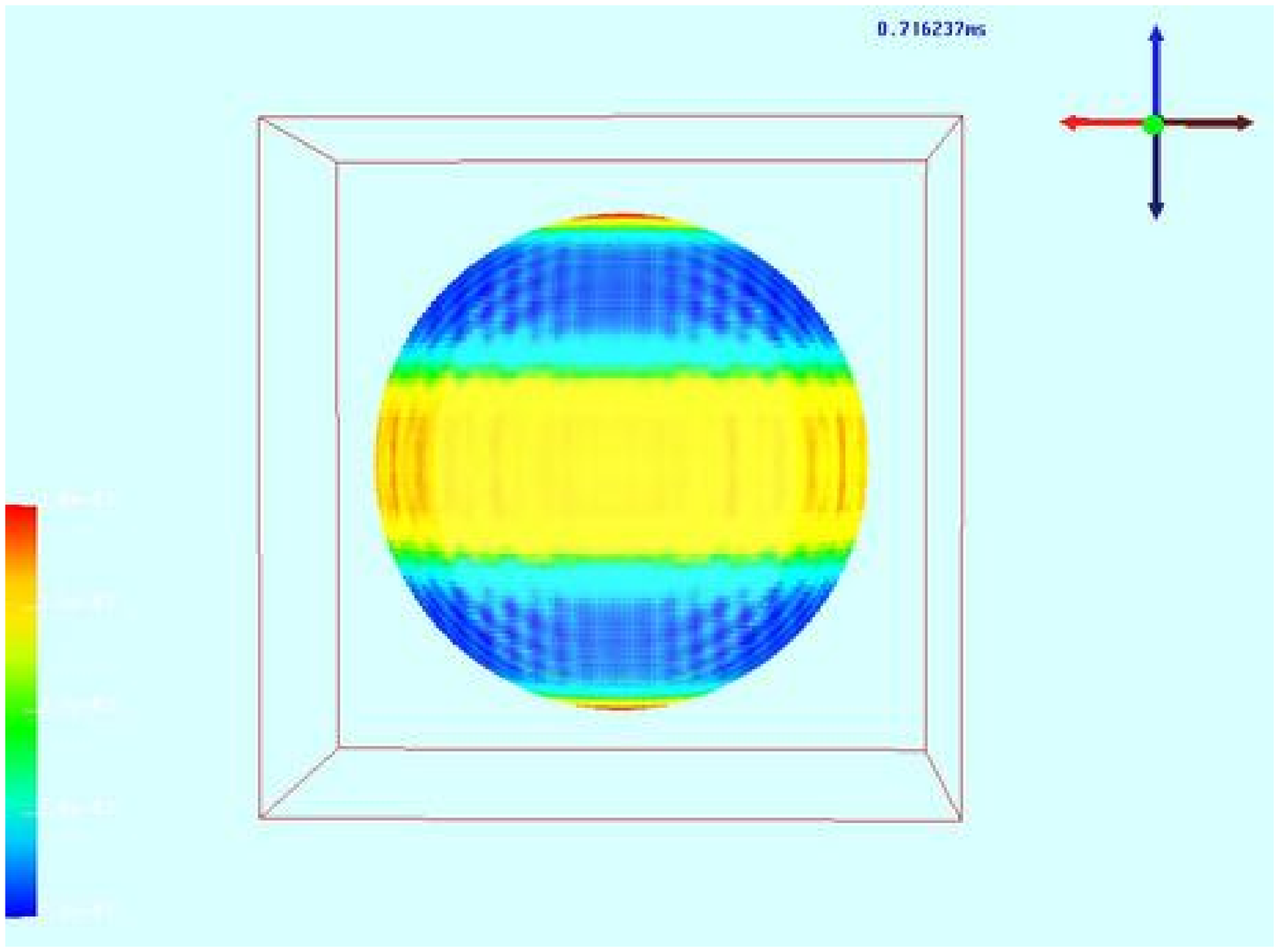}}
        \scalebox{0.30}{\includegraphics{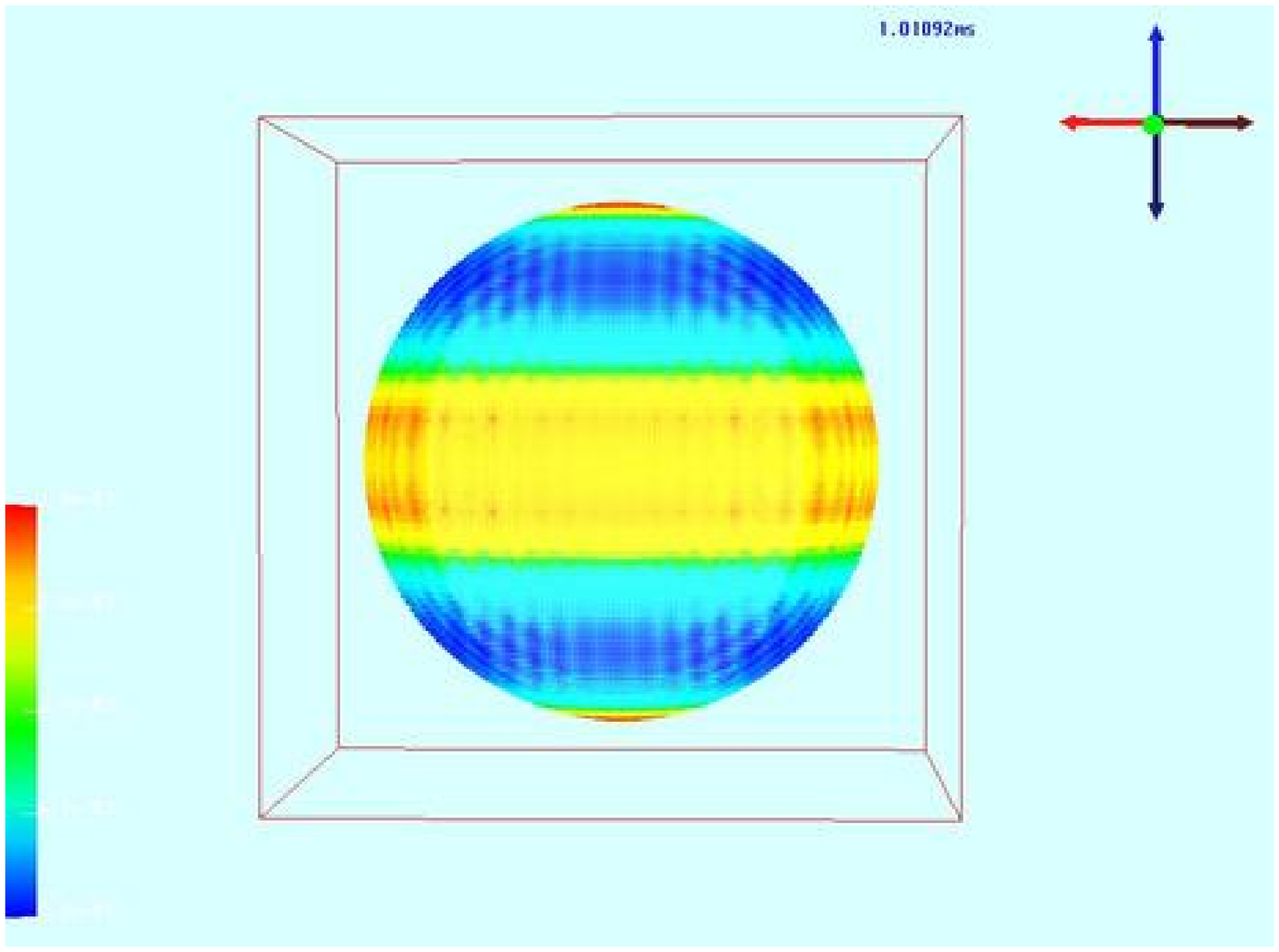}
                       \includegraphics{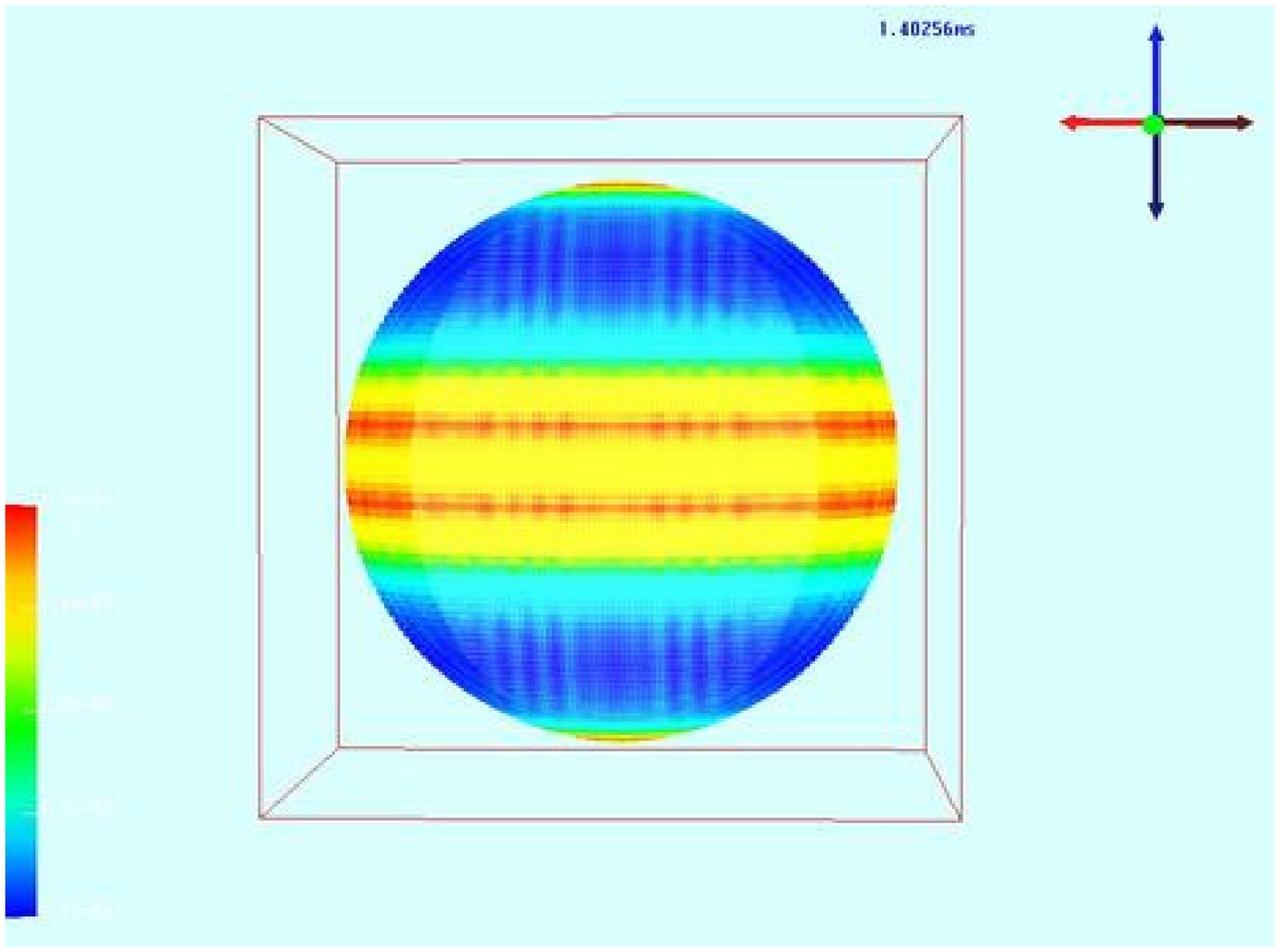}}
        \scalebox{0.30}{\includegraphics{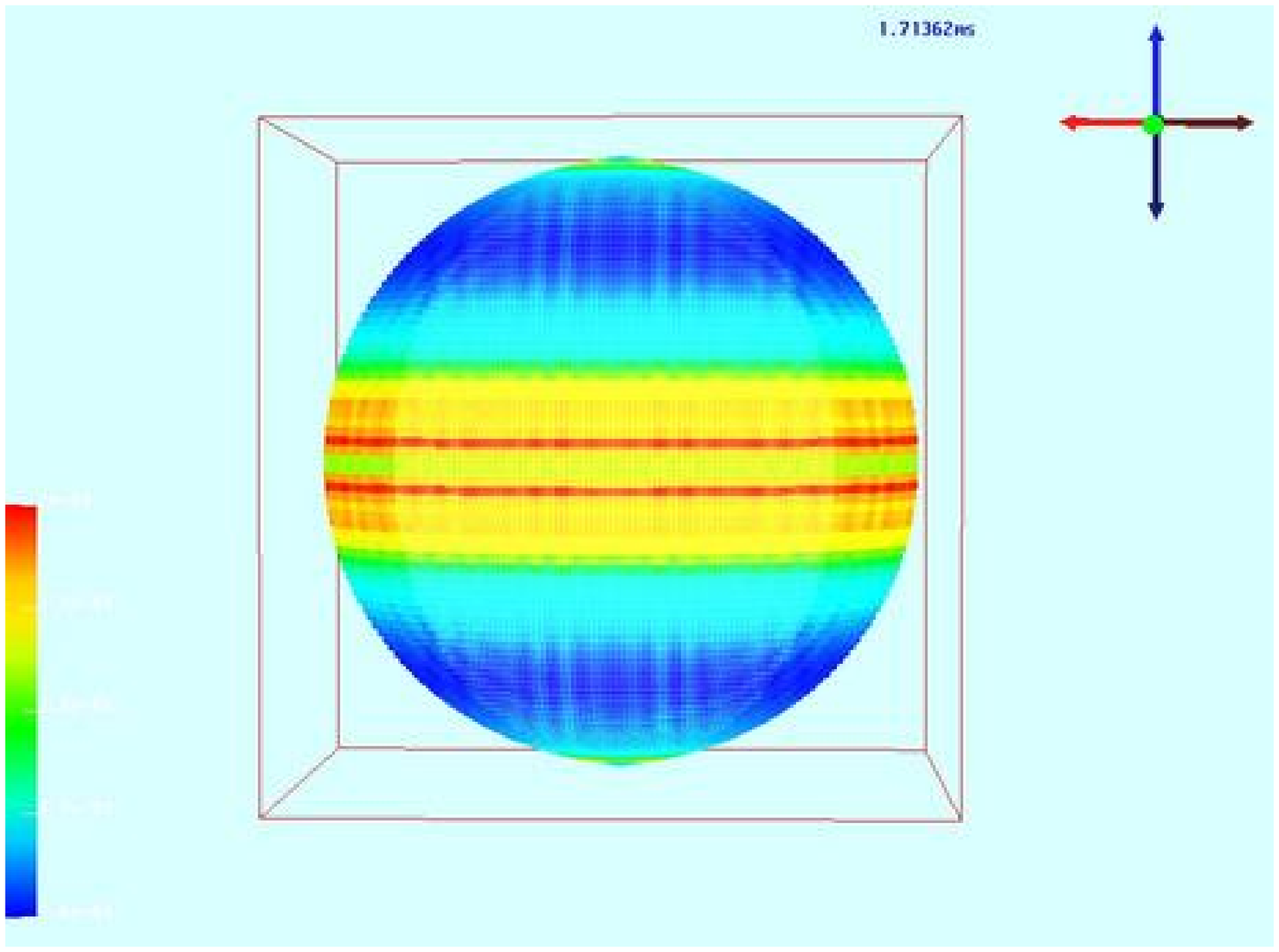}
                       \includegraphics{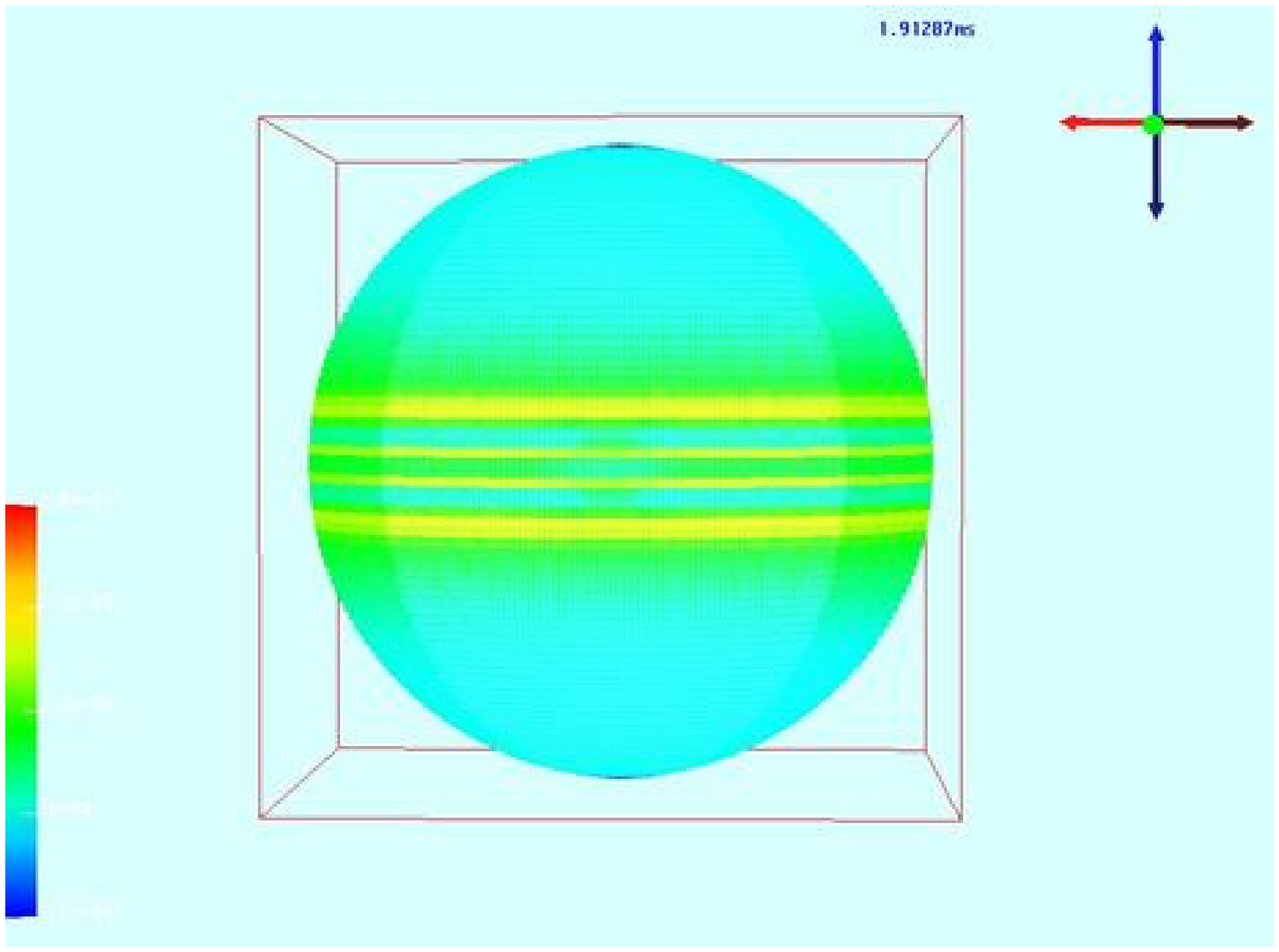}}
      \end{center}
      \caption{Deformation of inflating spherical shell.}
      \label{fig:sphShell}
    \end{figure}
    The particles in the figure are colored on the basis of the equivalent
    stress.  Though there is some difference between the values at different
    latitudes in the sphere, the equivalent stress is quite uniform in the
    shell.  The variation can be reduced using the implicit material point 
    method~\cite{Guilkey03}.
  \end{enumerate}

  \section{Summary}
  A shell formulation has been developed and implemented for the explicit 
  time stepping material point method based on the work of 
  Lewis et al.~\cite{Lewis98}.  Three different shell geometries and loading 
  conditions have been tested.  The results indicate that the stiff nature
  of the equation for rotational inertia may require the use of an implicit
  time stepping scheme for shell materials.

  \section*{Acknowledgments}
  This research was supported by a grant from ATK-Thiokol Propulsion. 

  \bibliographystyle{unsrt}
  \bibliography{../mybiblio.bib}

\end{document}